\documentclass[12pt]{article}

\usepackage{amssymb,amsfonts,amsmath,epsfig}
\usepackage{graphicx} 
\usepackage{indentfirst}
\usepackage{multirow}
\usepackage{color}

\parskip=\medskipamount

\usepackage[normalem]{ulem}

\renewcommand\sout{\bgroup \color{red} \ULdepth=-.5ex \ULset}

\newcommand{\sect}[1]{\setcounter{equation}{0}\section{#1}}

\newcommand{\be}{\begin{equation}}
\newcommand{\ee}{\end{equation}}
\newcommand{\bea}{\begin{eqnarray}}
\newcommand{\eea}{\end{eqnarray}}

\def\gkk{g^{\rm (1)}_{\rm KK}}

\begin{document}

\begin{titlepage}

\begin{flushright}
OCHA-PP-316
\end{flushright}
\vspace{2mm}

\begin{center}
{\Large \bf  Kaluza-Klein gluon searches using the three-$b$-jet decay channel 
 at the Large Hadron Collider} 
\end{center}

\begin{center}
%{\large  
{\bf
Masato Arai$^\dagger$$^\sharp$\footnote{masato.arai AT fukushima-nct.ac.jp},
Gi-Chol Cho$^\ddagger$\footnote{cho.gichol AT ocha.ac.jp}, and
Karel Smolek$^\sharp$\footnote{karel.smolek AT utef.cvut.cz}
} \\
\vspace{3mm}
{\it $^\dagger$Fukushima National College of Technology,
Fukushima, 970-8034, Japan} \\
{\it $^\ddagger$Department of Physics, Ochanomizu University, Tokyo 112-8610, Japan}\\
{\it $^\sharp$Institute of Experimental and Applied Physics,
Czech Technical University in Prague, 
Horsk\' a 3a/22, 128 00 Prague 2, Czech Republic}
\\

\vspace{2mm}

\end{center}
\vspace{5mm}

\begin{abstract}
\baselineskip=14pt
\noindent
We study observability of a Kaluza-Klein (KK) excitation of a gluon in 
a five-dimensional model with a warped geometry at the Large Hadron 
 Collider. In this model, the Standard Model fields reside in the bulk 
 and the third generation quarks couple to the KK gluon strongly. 
We focus on the processes including three $b$-quarks as a final state where 
the first KK gluon propagates as an intermediate state. We evaluate a 
 significance of those processes by taking account of kinematical cuts
 and a detector efficiency at the Large Hadron Collider and find that 
 the significance is lager than 5$\sigma$ with the integrated luminosity 
 of 10~(100)~fb${}^{-1}$ for a certain range of parameters of the model.  
\end{abstract}
\end{titlepage}

\newpage
\setcounter{page}{1}
\renewcommand{\thefootnote}{\arabic{footnote}}
\setcounter{footnote}{0}

\tableofcontents{}
\vspace{1cm}
\bigskip\hrule

%Masato130331
\sect{Introduction}
A possibility that our world is embedded into a higher dimensional 
space-time has been intensively studied. 
A strong motivation to study such a model is to solve the hierarchy 
problem between the electroweak scale and the Planck scale. 
The warped extra-dimensional model proposed by Randall and Sundrum (RS) 
\cite{RS} is one of promising candidates to explain such a large
hierarchy. 
This is a five-dimensional model where one extra-dimension is 
compactified on the orbifold $S^1/{\bf Z}_2$. There are two 3-branes 
located at different positions on this orbifold with opposite brane
tensions. 
One of the branes is called the visible brane where the Standard Model
(SM) fields localize and the other is called the hidden brane. 
Only a graviton can propagate into the whole space-time. 
With this setup, it is shown that an effective mass scale on the visible 
brane can be the electroweak scale as a consequence of the warped 
geometry, which is a natural solution of the hierarchy problem. 
%%% paragraph

%%% paragraph
Although in the original RS model all the SM fields localize on the 
brane, in order to solve the hierarchy problem, it is sufficient to
localize only the Higgs fields on the visible brane. It is allowed the 
other SM fields to propagate into the bulk together with graviton. 
We call this model the bulk RS model. 
Phenomenological aspects of this model have been investigated (for 
instance, see \cite{Davoudiasl:1999tf,Davoudiasl:2000wi}). 
A generic feature of the bulk RS model is that Kaluza-Klein (KK) 
excitations of the SM fields appear since the fermions and the gauge 
fields propagate into the bulk. 
They are signals of new physics if it is observed. In particular, at the 
Large Hadron Collider (LHC), one of such promising signals is the KK excitations
of the gluon since colored particles  
are dominantly produced at the LHC. 
%%% paragraph

%%% paragraph
The KK masses of the gluons are, however, severely constrained by 
tree-level flavor changing neutral currents (FCNC).  
Processes mediated by the KK excitations of the gluons can yield effects 
of large FCNC, which are constrained by low energy measurements of 
neutral mesons such as the $K^0-\bar{K^0}$ mixing. 
Contributions of KK gauge bosons to a CP violating parameter
$\epsilon_K$ in that process constrain the KK gauge boson mass to be 
$M_{KK}> 20$~TeV.  
However, it should be noted that these masses can be lowered by choosing
the bulk masses of fermions appropriately or assuming some flavor
structures of the quark and lepton sectors (for a convenient summary of
this issue, see \cite{JuWe}). 
%%% paragraph

%%% paragraph
Another constraint comes from the electroweak precision tests because of
the large contributions of the KK gauge bosons to the oblique
parameters. 
For instance, considering the model that only the gauge bosons propagate
into the bulk while the fermions are on the visible brane, the bound of 
the KK masses is $25-30$ TeV \cite{Davoudiasl:1999tf}. 
It corresponds to the typical scale of the RS model $\Lambda_\pi\sim
100$ TeV, which leads to the undesirable little hierarchy between the
electroweak scale and the lightest nonzero KK mass. 
It is, however, possible to lower the KK masses to be ${\cal O}$(TeV) by 
introducing the custodial symmetry \cite{Agashe:2003zs} or by placing
the SM fermions in the bulk \cite{HePeRi}.  
Even it is shown that models without the custodial symmetry make
possible the KK gauge boson mass lower to ${\cal O}$(TeV). 
As is seen above, the bound for the KK gauge boson mass is 
model-dependent and therefore it is worth examining observability of the 
KK gauge boson mass with a few TeV at the LHC. 
%%% paragraph

%%% paragraph
The KK gluon search in the bulk RS model has been studied in the 
processes including top quarks as a final state.  
For instance, in Refs. \cite{GuMaSr1,LiRaWa,Chang:2008vx,JuWe}, the
$t\bar{t}$ final state from the decay of 
$g_{\rm KK}^{(1)}\rightarrow t\bar{t}$ has been used as a probe of 
the KK gluon, where $g_{\rm KK}^{(1)}$ is the first KK excitation of 
a gluon. 
The associate production of $g_{\rm KK}^{(1)}$ decaying to $t\bar{t}$
has been studied in Ref. \cite{GuMaSr2}. 
Experimental bounds have been obtained by using the decay channel 
$g_{\rm KK}^{(1)}\rightarrow t\bar{t}$. 
The experimental lower bound on the mass of $g_{\rm KK}^{(1)}$ is of 
order $1$~TeV in both the Tevatron \cite{CDF} and the LHC \cite{CMS, ATLAS}. 
%%% paragraph

%%% paragraph
The purpose of our work is to study a sensitivity of the first KK 
excitation of a gluon in the bulk RS model at the LHC by using the $b$
quark final state only, in particular, the process $pp \rightarrow 3b$. 
In our previous paper \cite{ArChSmYo}, we studied a sensitivity to 
observe the KK gluon at the LHC through the decay channel 
$g_{\rm KK}^{(1)}\rightarrow b\bar{b}$ for various choices of couplings
of the KK gluons to the SM quarks.  
We found that the KK gluon with the mass up to $1.4$~TeV would be 
observable with a significance $5\sigma$ in the choice of parameters for
integrated luminosity of $100$ fb${}^{-1}$. 
In this paper, we consider the case that the KK gluon strongly couples to the 
right-handed top-quark $t_R$ since it is naively expected that such an 
interaction does not affect electroweak and flavor processes 
\cite{GuMaSr1,LiRaWa,JuWe,AgBeKrPeVi}.  
In addition, we choose the couplings 
of the KK gluon to the 
right-handed bottom quark
to be as large as the coupling of the KK gluon to the $t_R$ and to be smaller than it.
In the literature, it is chosen to be negligibly suppressed since it
contributes to FCNC. 
However, there are various variants of the Yukawa sector in the  bulk RS
model (e.g., Ref. \cite{Chang:2008vx}). 
Therefore, we leave the coupling of the KK gluon to the left- and
right-handed $b$ quark as phenomenological parameters. 
Detailed choices will be found in the main body.  
%%% paragraph

%%% paragraph
Organization of this paper is as follows. In Sec. 2, we briefly review
the bulk RS model with the SM bulk fermions and the gauge bosons. 
Numerical results are shown in Sec.~3. Sec. 4 is devoted to summary. 

%%%%%%%%%%%%%%%%%%%%%%%%%%%
\sect{Model}
In this section, we briefly review the bulk RS model focusing on the 
interactions of the KK gluon (we follow the notation of 
Ref. \cite{GhPo}).  
The RS model is a five-dimensional model with a warped fifth dimension.  
The space-time metric is given by 
\begin{eqnarray}
 ds^2=G_{MN}dx^M dx^N =
e^{-2\sigma}\eta_{\mu\nu}dx^\mu dx^\nu+dy^2,
	\label{metric}
\end{eqnarray}
where $\eta_{\mu\nu}={\rm diag}(-1,+1,+1,+1)$ is the Minkowski metric 
and $\sigma=k|y|$. 
Throughout this paper, the roman index runs from 0 to 4 while the Greek 
index runs from 0 to 3. 
The curvature of the $AdS_5$ is determined by a dimensionful parameter 
$k$.  
Two 3-branes are located at the fixed points of the orbifold 
$S^1/{\bf Z}_2$ with a radius $r_c$ at $y=0$ and $\pi r_c$, which are 
called the hidden brane and the visible brane, respectively. 
With this setup, the effective mass scale at the visible brane 
is given as $M_P e^{-k\pi r_c}$, where $M_P$ is the four-dimensional 
Planck scale. 
The effective mass reproduces the electroweak scale when $k r_c \simeq
12$ so that the hierarchy between the electroweak scale and the Planck
scale is stabilized without introducing a serious fine-tuning. 
In the following, we assume that only the SM Higgs field localizes in 
the visible brane while the other fields propagate into the 
five-dimensional bulk.  
%%% paragraph

%%% paragraph
The gauge invariant action of a gauge field $A_M^a$ and a fermion 
$\Psi$ is given by  
\begin{eqnarray}
S_5=-\int d^4x \int dy \sqrt{-G}\left[{1 \over 4}F_{MN}^aF^{MNa} 
+i\bar{\Psi}\gamma^M D_M\Psi+i m_\Psi \bar{\Psi}\Psi \right], 
\label{action}
\end{eqnarray}
where $G=\det{(G_{MN})}$, $F_{MN}^a$ denotes the field strength of a 
gauge field $A_M^a$ and $D_M$ is a covariant derivative. 
The index $a=1,\cdots, {\rm dim}({\cal G})$ is a gauge index, where 
${\rm dim}({\cal G})$ is a dimension of a gauge group ${\cal G}$. 
In the following, we choose the color gauge group ${\rm SU}(3)_C$ 
relevant to dynamics of the gluon.  
The five-dimensional fermion mass parameter $m_\Psi$ is given as 
\begin{eqnarray}
  m_{\Psi}=ck\epsilon(y),
\end{eqnarray}
where $c$ is an arbitrary dimensionless parameter, and $\epsilon(y)$ is 
$+1$ for $y>0$ while $-1$ for $y<0$ to make the mass term to be 
${\bf Z}_2$ even.  
%%% paragraph

%%% paragraph
The KK decomposition of a ``gluon'' $A^a_M(x,y)$ and $\Psi(x,y)$ is 
given as\footnote{We adopt a gauge choice $A_4=0$.}
\begin{eqnarray}
 \Phi(x^\mu,y)
={1 \over \sqrt{2\pi r_c}}\sum_{n=0}^\infty \Phi^{(n)}(x^\mu)f_n(y), 
\label{modeE}
\end{eqnarray}
where $\Phi=\{A^a_\mu,\Psi\}$ and 
$\Phi^{(n)}(x^\mu)=\{A^{a(n)}_\mu(x^\mu),\psi^{(n)}(x^\mu)\}$.  
The mode functions $f_n(y)$ satisfy the orthonormal condition 
\begin{eqnarray}
 {1 \over 2\pi r_c}\int_{-\pi r_c}^{\pi r_c}dy 
e^{(2-s)\sigma}f_n(y)f_m(y)=\delta_{nm}, 
\label{on}
\end{eqnarray}
with $s=2,1$ for $A^a_\mu, \Psi$.
The explicit form of the function $f_n$ for the zero ($n=0$) and 
$n(\neq 0)$ modes are given as follows 
\begin{eqnarray}
 f_0 &=& 1, \label{sol1}
\\
 f_n &=& \frac{e^{\sigma}}{N_n}\left[
J_1\left(\frac{m_n}{k}e^{\sigma}\right)
+ b_1(m_1)Y_1\left(\frac{m_n}{k}e^{\sigma}\right)\right], \label{sol2}
\end{eqnarray}
where $J_1$ and $Y_1$ are the standard Bessel functions of the first and 
second kind, $N_n$ is a normalization factor, $b_1(m_n)$ is the
constant, and $m_n$ is the mass of the $n$-th KK mode. 
%%% paragraph

%%% paragraph
Substituting (\ref{modeE}) with 
(\ref{sol1}) and (\ref{sol2}) 
into (\ref{action})
and integrating the $y$ direction, we find the interactions of 
the gluon and fermions with an arbitrary KK index $n$. 
The four-dimensional SU(3)$_C$ gauge coupling $g_4$ is obtained from
three gluon interactions in (\ref{action}) by integrating out the mode
functions as 
\begin{eqnarray}
 g_4 = g_5\int^{\pi r_c}_{-\pi r_c} dy \frac{f_0^3}{(2\pi r_c)^{3/2}}
=\frac{g_5}{\sqrt{2\pi r_c}}. 
\end{eqnarray}
The condition (\ref{on}) tells us that a KK gluon 
$g_{\rm KK}^{(n)}~(n\neq 0)$   
does not couple to a zero-mode gluon pair, i.e., there is no  
$g_{\rm KK}^{(n)}$-$g$-$g$ vertex,
where $g_{\rm KK}^{(n)}$ and $g$ denote the $n$-th excitations of the KK gluon and the four-dimensional gluon, respectively.
It is also easy to see that the couplings of 
$g_{\rm KK}^{(n)}$-$g_{\rm KK}^{(n)}$-$g$ and $g_{\rm KK}^{(n)}$-$g_{\rm
KK}^{(n)}$-$g$-$g$ 
vertices are identical to the $g$-$g$-$g$ and $g$-$g$-$g$-$g$ vertices, 
respectively~\cite{Allanach:2009vz}. 
The effective coupling of the interaction vertex of the $n$-th KK gluon 
and the zero mode fermion pair is 
\begin{eqnarray}
 g^{(n)}=g_4{1-2c \over e^{(1-2c)\pi kr_c}-1}
{k \over N_n}
 \left[J_1\left({m_n \over k}e^\sigma \right)+b_1(m_n)Y_1\left({m_n \over k}e^\sigma \right)\right]. \label{coupling}
\end{eqnarray}
More detailed discussion is given in, for example,
Refs.~\cite{GhPo,Allanach:2009vz}. 
%%% paragraph

%%% paragraph
%%
We consider the following scenarios with various values of couplings.
\begin{eqnarray}
&&{\rm (1)}: {g_{Q_3}^{(1)} \over g_4}={g_{t}^{(1)} \over g_4}={g_{b}^{(1)} \over g_4}=4, \quad {g_{\rm light}^{(1)} \over g_4}=0, \label{couplings1}\\
&&{\rm (2)}: {g_{Q_3}^{(1)} \over g_4}=1, \quad {g_{t}^{(1)} \over g_4}={g_{b}^{(1)} \over g_4}=4, \quad {g_{\rm light}^{(1)} \over g_4}=0, \label{couplings2}\\
&&{\rm (3)}: {g_{Q_3}^{(1)} \over g_4}=1, \quad {g_{t}^{(1)} \over g_4}=4,\quad {g_{b}^{(1)} \over g_4}=1,\quad{g_{\rm light}^{(1)} \over g_4}=0, \label{couplings3}\\
&&{\rm (4)}: {g_{Q_3}^{(1)} \over g_4}=1, \quad {g_{t}^{(1)} \over g_4}=4,\quad {g_{b}^{(1)} \over g_4}={g_{\rm light}^{(1)} \over g_4}=0. \label{couplings4}
\label{couplings}
\end{eqnarray}
where $Q_3$ is the third generation of the left-handed quark, $t,b$ are
the right-handed top and bottom quarks and ``light'' means the quarks of 
the first two generations. In (1), couplings of all the quarks of the
3rd generation to the KK gluon is strong while a coupling between the KK
gluon and the light quarks is vanishing. The latter choice is motivated
by the constraints coming from the FCNC and the electroweak precision 
measurement.  In (2),  the KK gluon strongly couples to the right-handed
quarks only. The coupling to the left-handed quark is comparable to the
QCD coupling $g_4$. This choice has been studied to analyze a decay of the
KK gluon to top and anti-top quarks \cite{LiRaWa}
\footnote{
Note that the strong coupling of the right-handed $b$ quark to the KK
gluon may give a sizable shift of the $Zbb$ vertex in 1-loop level. 
Such an effect is, however, marginal even for $g_b^{(1)}/g_4=6$ when the KK
gluon mass is larger than 500 GeV \cite{Cho:2009rj}. Therefore our
choice of the couplings in (1) and (2) does not affect the $Zbb$ vertex 
since we take $M_{\gkk} \ge 1$ TeV in our study. 
}. 
In (3), the KK gluon coupling to the right-handed bottom quark 
is taken to be $1$ while the rest of the choice is the same as (2). This is motivated by 
the mass hierarchy between the top quark mass and the bottom quark,
which would affect the couplings through (\ref{coupling}).
In (4), the difference from (3) is to take the KK gluon coupling to the right-handed
bottom quark to be zero.
This choice comes from the flavor physics and the electroweak precision 
measurement \cite{JuWe}.
With the above choice of parameters, we perform the numerical analysis of the process 
$pp\xrightarrow{g_{\rm KK}^{(1)}} 3b$.

\sect{Numerical Analysis}

\begin{figure}[tb]
\begin{center}
\begin{eqnarray*}
 \begin{array}{cc}
  \epsfxsize=6cm
  \epsfbox{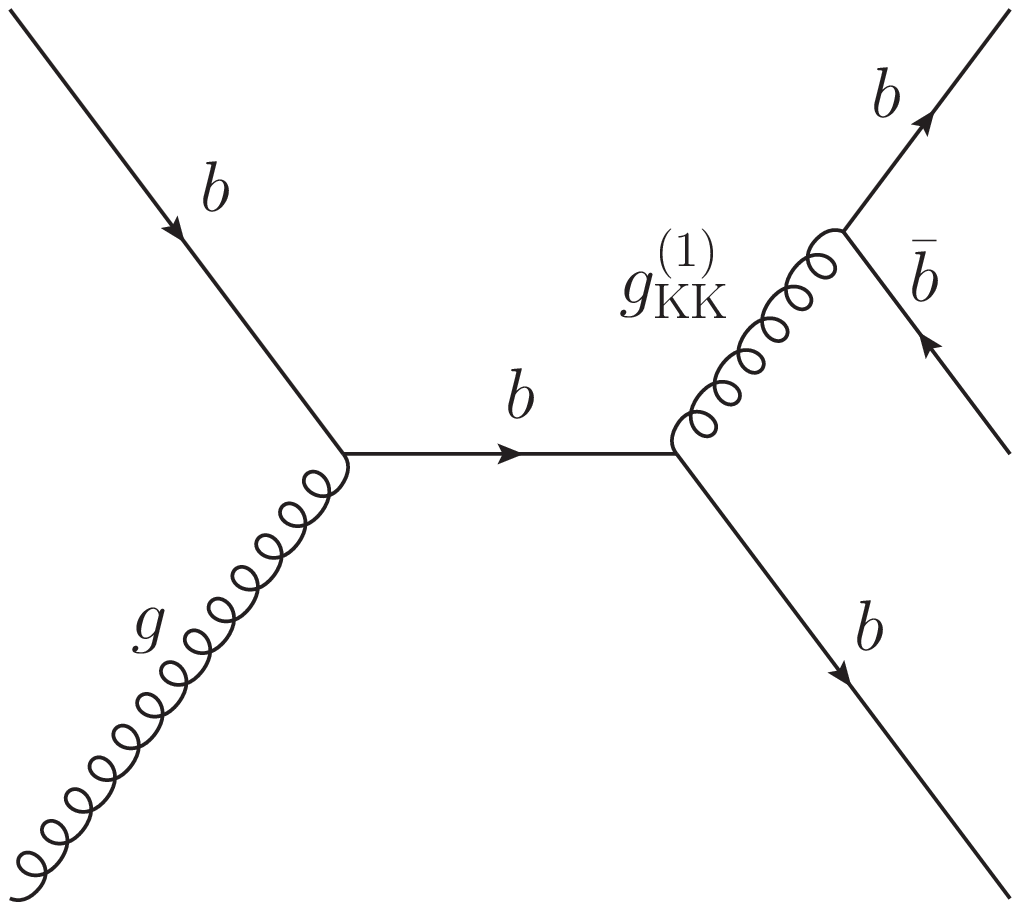}
  &
  \epsfxsize=6cm
  \epsfbox{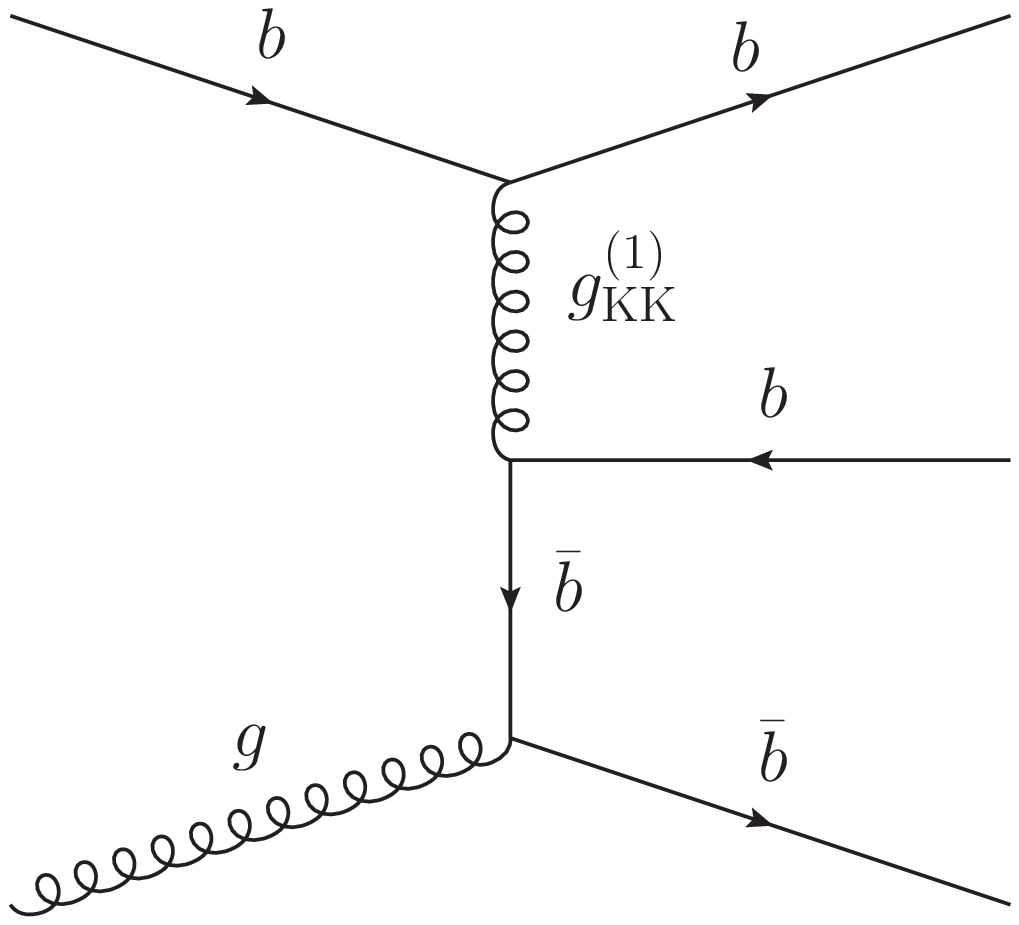} \\
  \epsfxsize=6cm
  \epsfbox{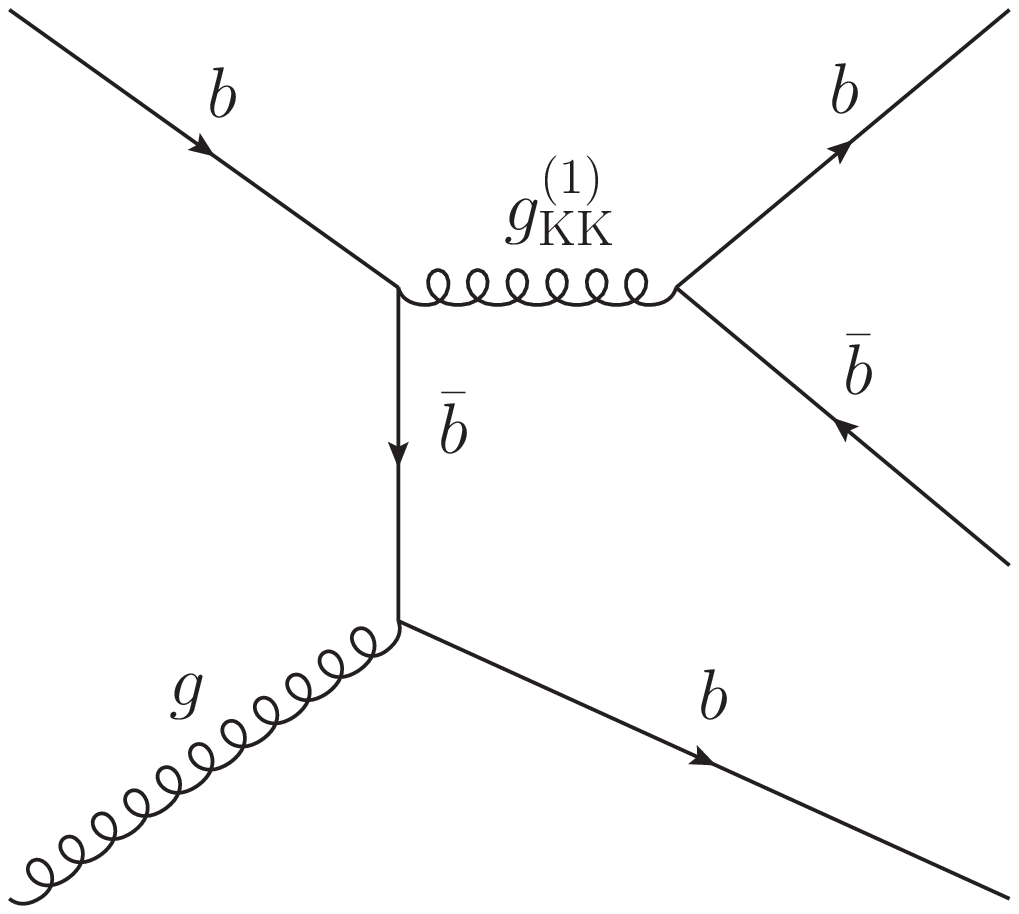}
  &
  \epsfxsize=6cm
  \epsfbox{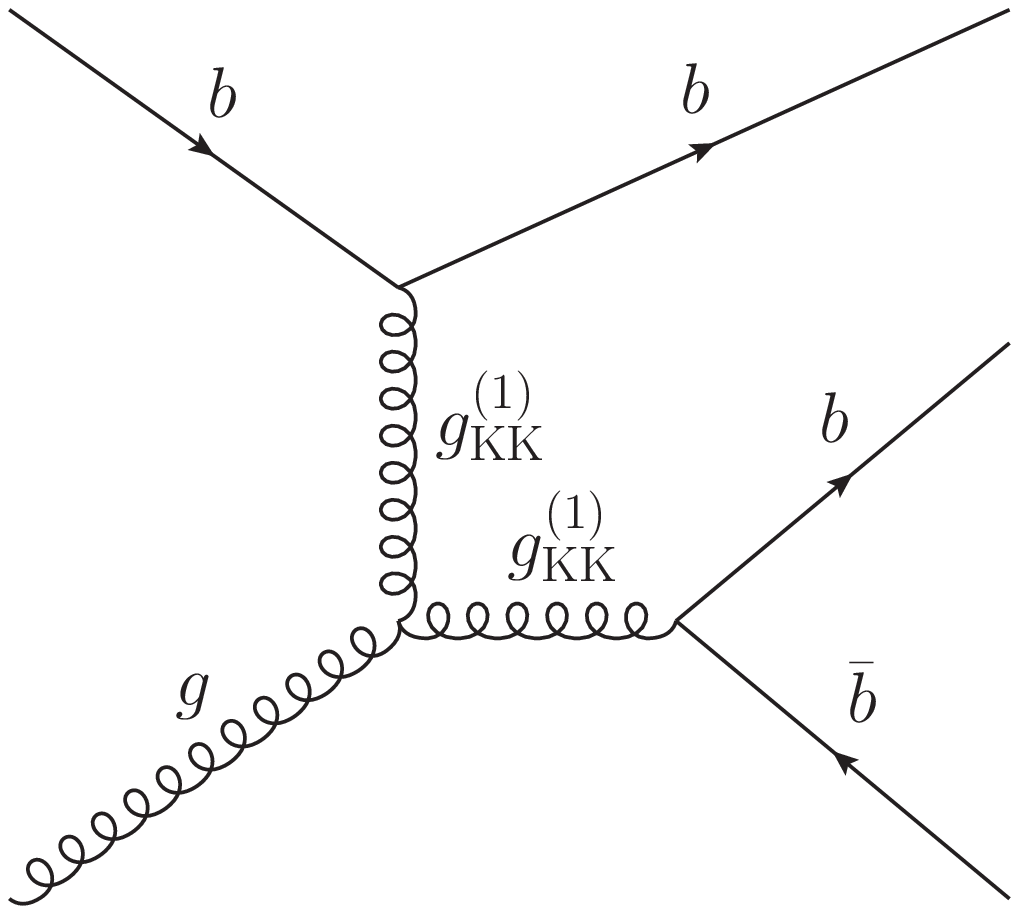} \\
 \end{array}
\end{eqnarray*}
\caption{
The representative Feynman diagrams describing the studied processes
 with three $b$-jets in the final state. 
The rest of the diagrams can be obtained by using one or more
transformations from the following list: interchanging of $b$ and
$\bar{b}$ in a whole diagram, changing of order of initial partons,
changing of order of final partons. 
}
  \label{Diagrams}
\end{center}
\end{figure}

We study effects of the KK gluon  
predicted by the presented model in the three $b$ final states ($bb\bar{b}$ and $b\bar{b}\bar{b}$) at the
LHC. Four representative Feynman diagrams describing the studied
processes in the leading order are visualized at the
Fig. \ref{Diagrams}. 
The rest of the diagrams can be obtained by using one or more
transformations from the following list: interchanging of $b$ and
$\bar{b}$ in a whole diagram, changing of order of initial partons,
changing of order of final partons. 
The presented diagrams manifest that studied final states can be produced only with the interaction of the initial $b$ quark and the gluon. 
Corresponding matrix elements for the signal process were obtained 
by
using MadGraph 5.1.5.7 \cite{MadGraph} where we implemented the KK gluon in MadGraph5 
with
FeynRules \cite{Christensen:2008py}.
The signal events on the parton level were simulated 
by 
using MadEvent package, a part of the MadGraph. We simulated possible background processes, initial/final state
radiations, hadronization and decays for both, the signal and background processes, 
by 
using Pythia 8.175 \cite{Pythia6,Pythia8}, the Monte-Carlo generator 
with  
leading-order 
expressions of matrix-elements. All samples were generated for $pp$
collisions at $\sqrt{s} = 14$ TeV 
by 
using the CTEQ6L1 parton distribution
functions \cite{CTEQ6L1}. The factorization scale $Q_F$ for the $2\rightarrow 3$ processes was chosen to  be equal to the geometric mean of the two smallest squared transverse masses of the three outgoing particles. The renormalization scale $Q_R$ was chosen to be equal to the geometric mean of the squared transverse masses of the three outgoing particles. 
To save time in the simulations of the background $ab\rightarrow cde$  processes, the phase space cuts $\hat{p}_{T}(c) > 50$~GeV, $\hat{p}_{T}(d) > 50$~GeV and $\hat{p}_{T}(e) > 50$~GeV on the transverse momentum of final particles in their center-of-mass system were applied.

For the simulation of the effects of a detector, we used Delphes 3.07 \cite{Delphes}, a framework for a fast simulation of a generic collider experiment. The fast simulation of the detector includes a tracking system, a magnetic field of a solenoidal magnet affecting tracks of charged particles, calorimeters and a muon system. The reconstructed kinematical values are smeared according to the settings of the detector simulation. For the jets reconstruction, Delphes uses the FastJet tool \cite{FastJet1,FastJet2} with several implemented jet algorithms. In our simulations, we used the data file with standard settings for the ATLAS detector, provided by the tool. We used the $k_T$ algorithm \cite{kT_algorithm} with a cone radius parameter $R = 0.7$. The $b$-tagging efficiency is assumed to be 40\%, independently on a transverse momentum and a pseudorapidity of a jet. A fake rate of the $b$-tagging algorithm is assumed to be 10\% for $c$-jets and 1\% for light- and gluon-jets. These settings for the $b$-tagging are standard for the ATLAS detector in the Delphes tool. No trigger inefficiencies are included in this analysis.

\begin{figure}[tb]
\begin{center}
  \epsfxsize=12cm
  \epsfbox{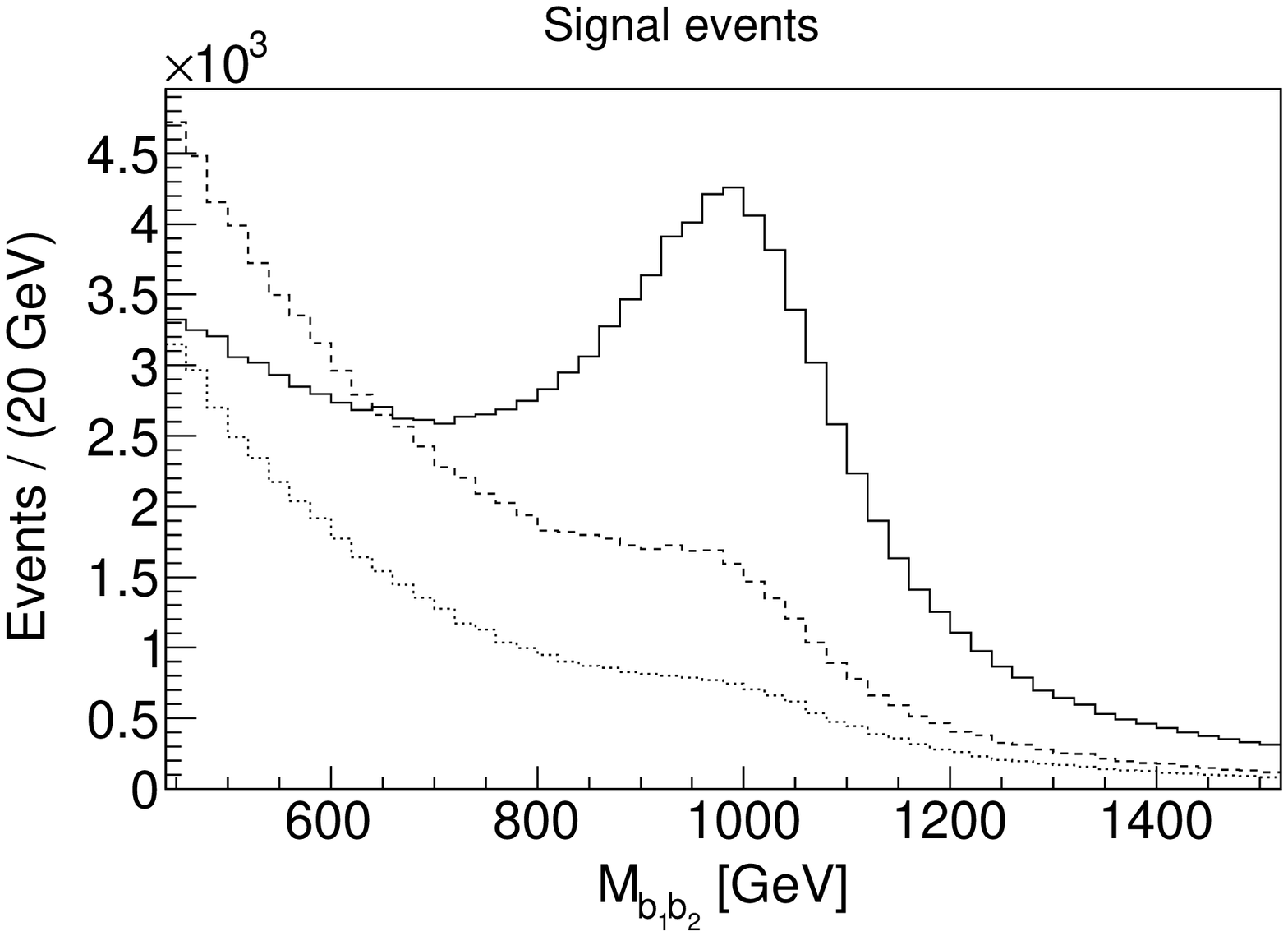} \\
\caption{
The invariant mass distribution of all three $b$ quark pairs for the signal process for $M_{\gkk} = 1$ TeV and the scenario (2) in (\ref{couplings2}). The full line corresponds to the pair with the two highest transverse momenta, the dashed line corresponds to the pair with the highest and lowest transverse momenta and the dotted line corresponds to the pair with the lowest transverse momenta. The number of events in the histogram is scaled to the integrated luminosity of 10 fb$^{-1}$ for $pp$ collisions at $\sqrt{s}=14$ TeV. 
}
\label{Mbbb}
\end{center}
\end{figure} 

\begin{figure}[tb]
\begin{center}
  \epsfxsize=12cm
  \epsfbox{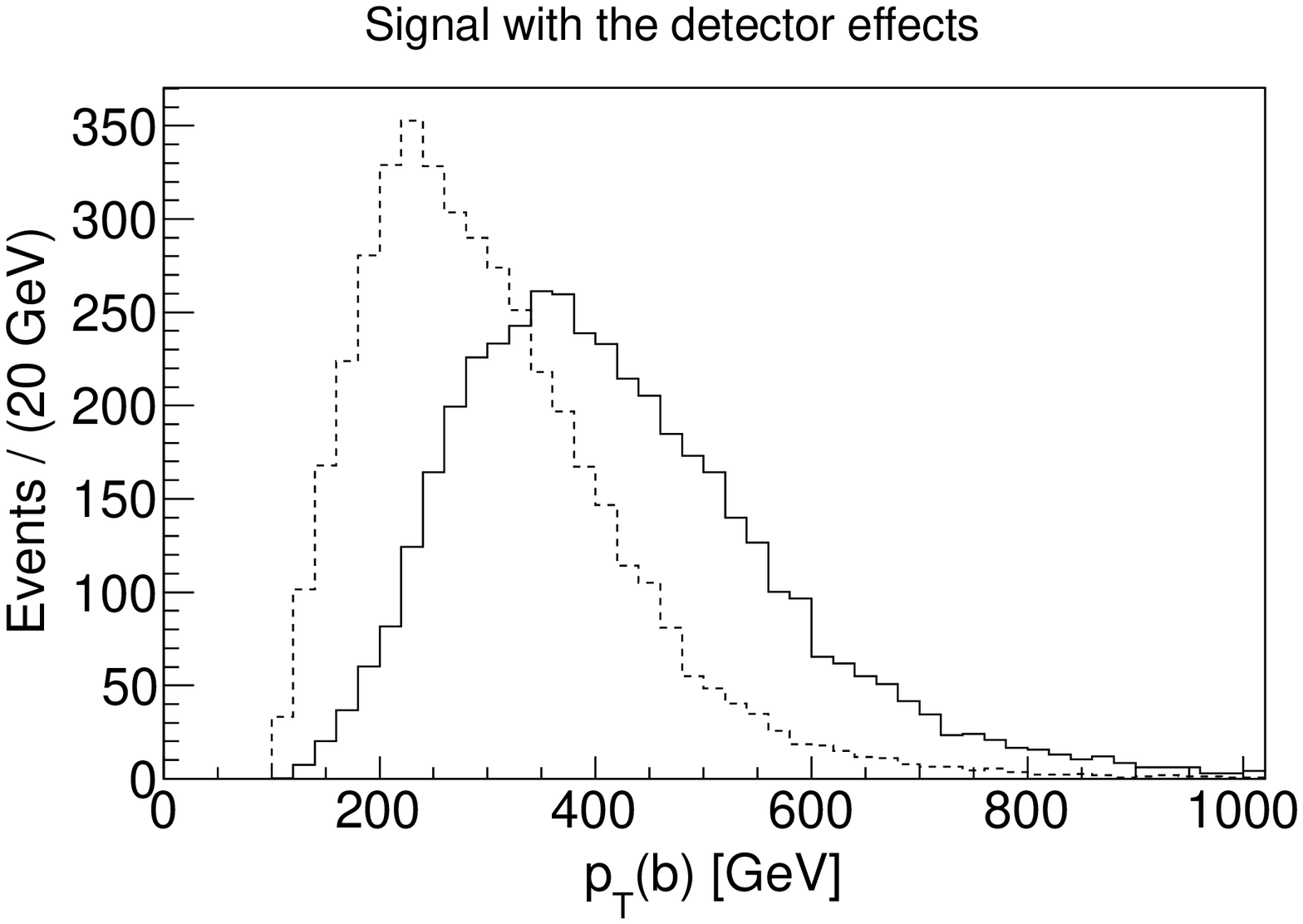} \\
\caption{
The transverse momentum distribution for $b$-jets selected 
in the scenario (2) in (\ref{couplings2}) with $M_{\gkk} = 1$~TeV
after applying of the Selection criterion 1 and $M_{b_1b_2}^{min}$ = 450 GeV for the signal events.  
The full line corresponds to the $b$-jet with the highest transverse momentum, and the dashed line corresponds to the $b$-jet with the second highest transverse momentum. The number of events in the histogram is scaled to the integrated luminosity of 10 fb$^{-1}$ for $pp$ collisions at $\sqrt{s}=14$~TeV. 
}
\label{Pt}
\end{center}
\end{figure} 

For the three $b$-jets final states, it is difficult to select two $b$-jets corresponding to the decay products of the KK gluon. 
In the Fig. \ref{Mbbb}, the distribution of the invariant mass for all three $b$ quark pairs for the signal process with $M_{\gkk} = 1$ TeV 
and the scenario (2) is plotted. 
The resonance associated with a KK gluon is most evident in the invariant
mass distribution of two jets with the highest transverse momenta in the invariant mass of two jets with the highest transverse momenta. 
Therefore, we assume this selection criterion:
\begin{description}
  \item[Selection criterion 1] The event must have exactly three $b$-tagged jets with the transverse momentum in the laboratory frame $p_T > 100$~GeV and the pseudorapidity $|\eta| < 2.5$. The invariant mass of two $b$-jets $b_1$, $b_2$ with the highest transverse momenta must fulfill the condition $M_{b_1b_2} > M_{b_1b_2}^{min}$  for a chosen $M_{b_1b_2}^{min}$.
\end{description}
Due to the $b$-tagging and sufficiently high $M_{b_1b_2}^{min}$, the criterion effectively suppresses all QCD background processes (including a top quark production with all decay channels). In our analysis, the $b$-tagging is used for jets with very high transverse momentum (hundreds of GeV), see Fig. \ref{Pt}. At present, most of $b$-tagging algorithms are optimized and intensively tested for a lower range of jet transverse momenta. Therefore, we studied possibility not to use 
the 
$b$-tagging and to rely on strict selection cuts on the transverse momentum and invariant mass of the detected jets, instead:
\begin{description}
  \item[Selection criterion 2] The event must have at least three jets with the transverse momentum in the laboratory frame 
  $p_T > 100$~GeV and the  pseudorapidity $|\eta| < 2.5$. The invariant mass of two jets $b_1$, $b_2$ with the highest transverse momenta must fulfill the condition $M_{b_1b_2} > M_{b_1b_2}^{min}$ for a chosen $M_{b_1b_2}^{min}$.
\end{description}

\begin{figure}[htb]
\begin{center}
\begin{eqnarray*}
 \begin{array}{cc}
  \epsfxsize=5.9cm
  \epsfbox{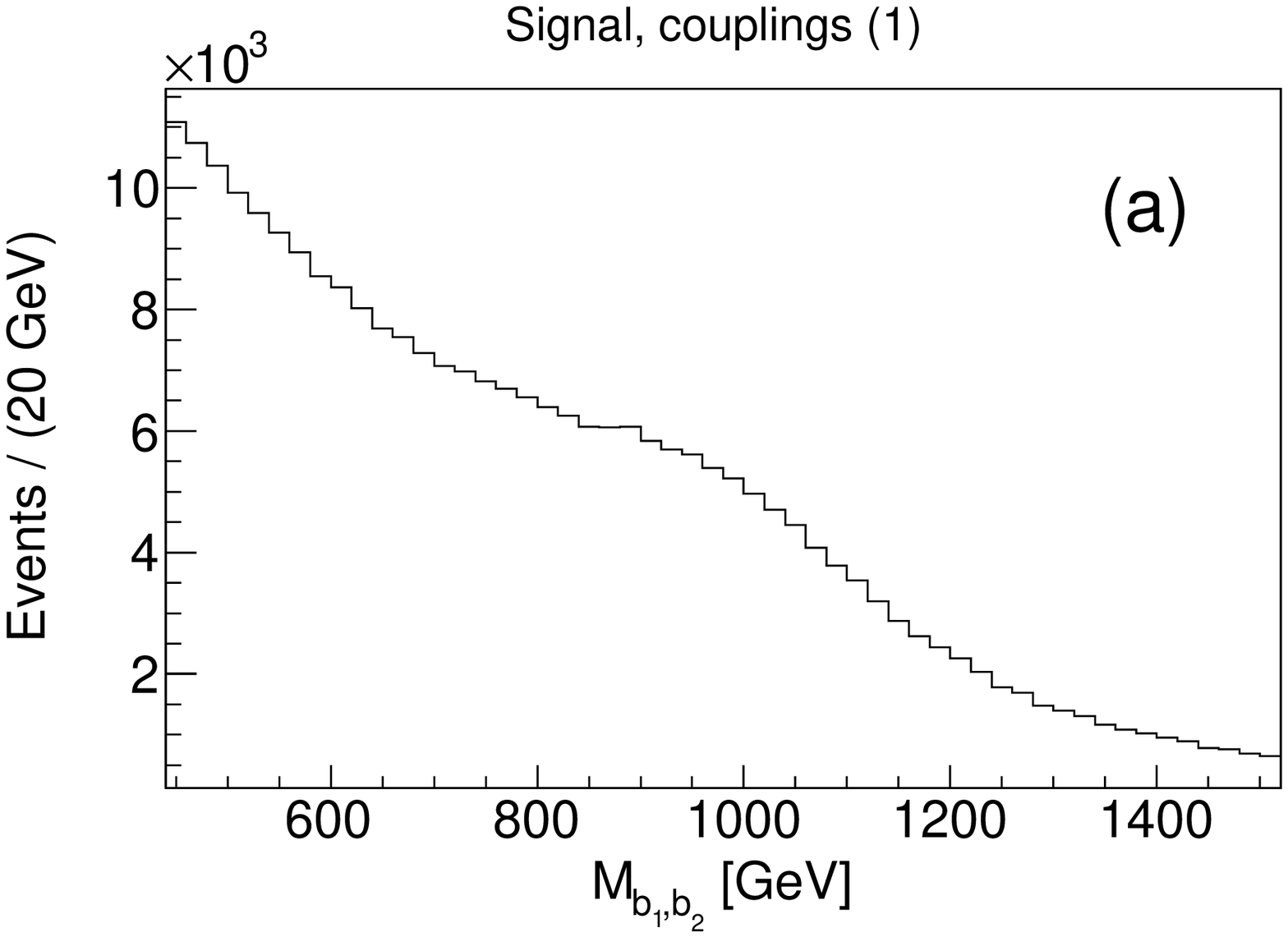}
  &
  \epsfxsize=5.9cm
  \epsfbox{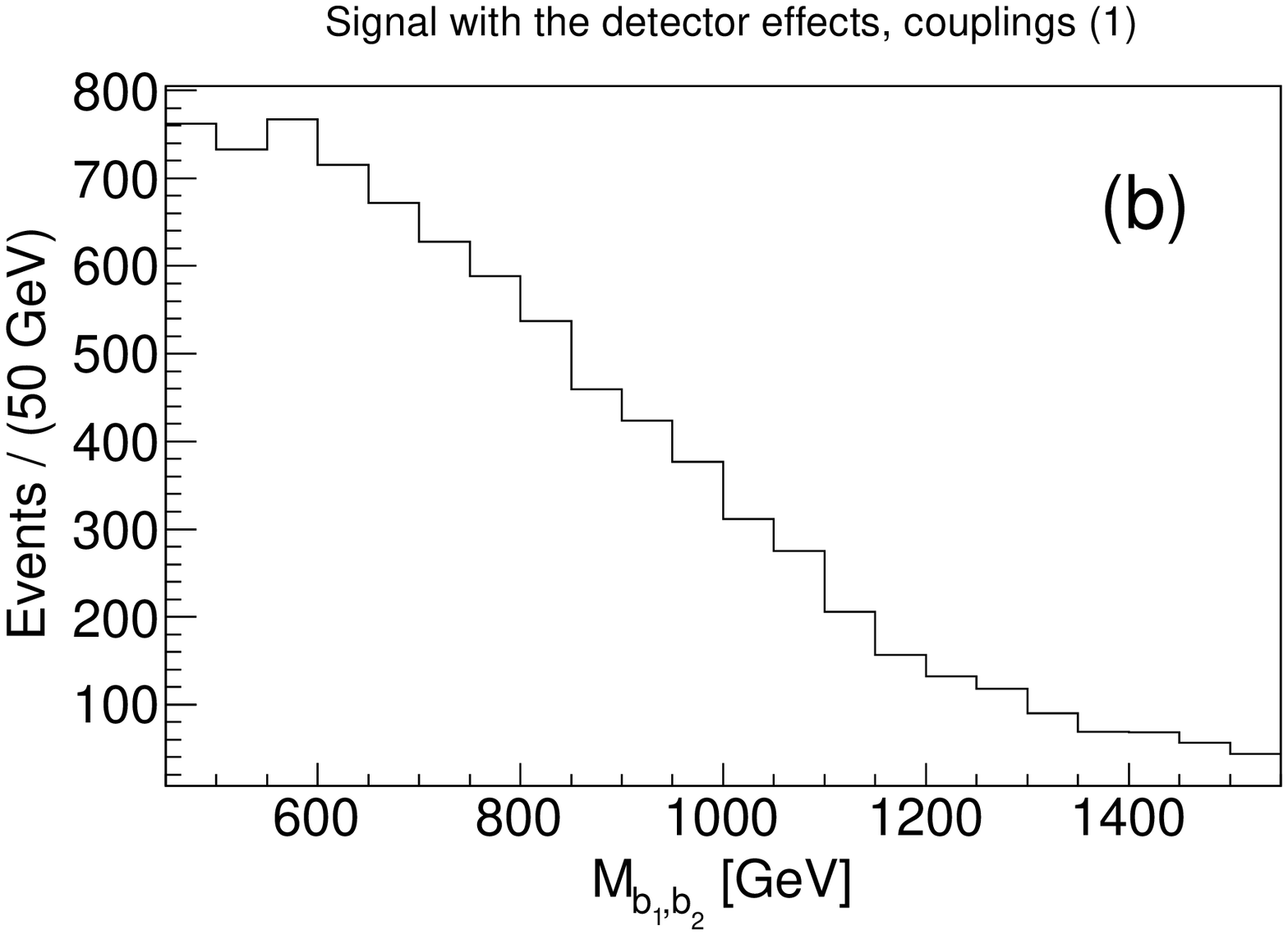} \\
  \epsfxsize=6cm
  \epsfbox{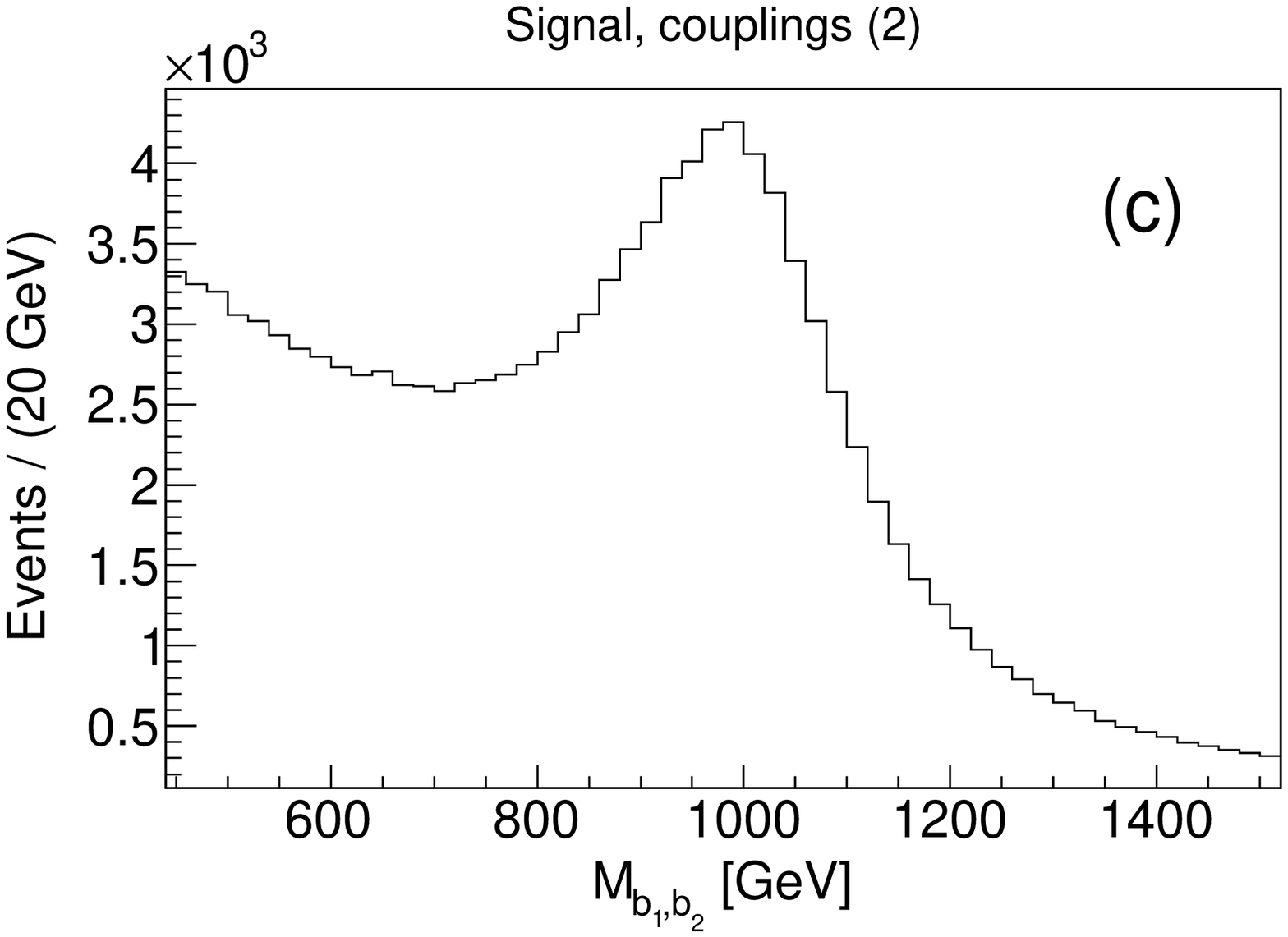}
  &
  \epsfxsize=5.9cm
  \epsfbox{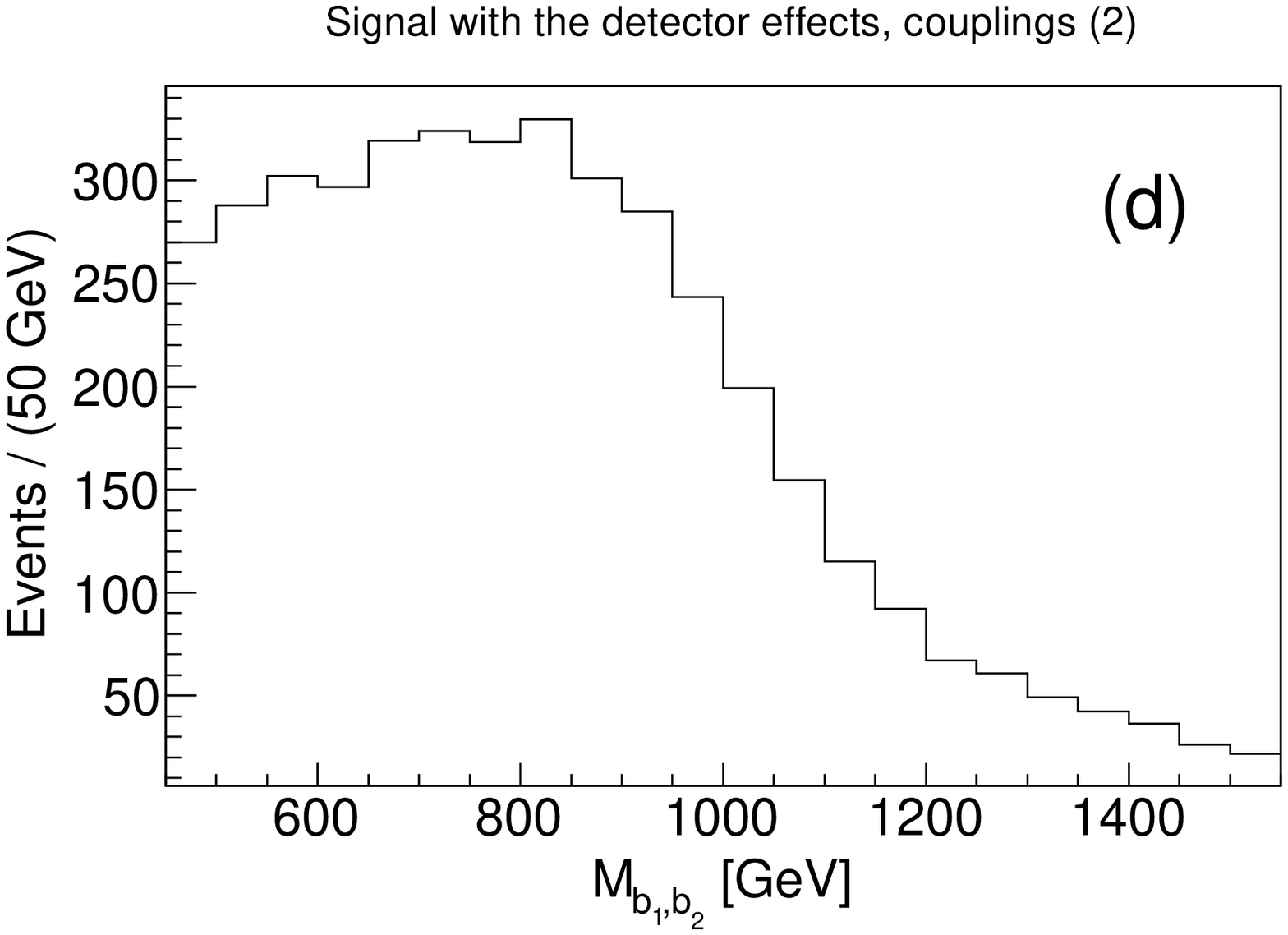} \\
  \epsfxsize=5.9cm
  \epsfbox{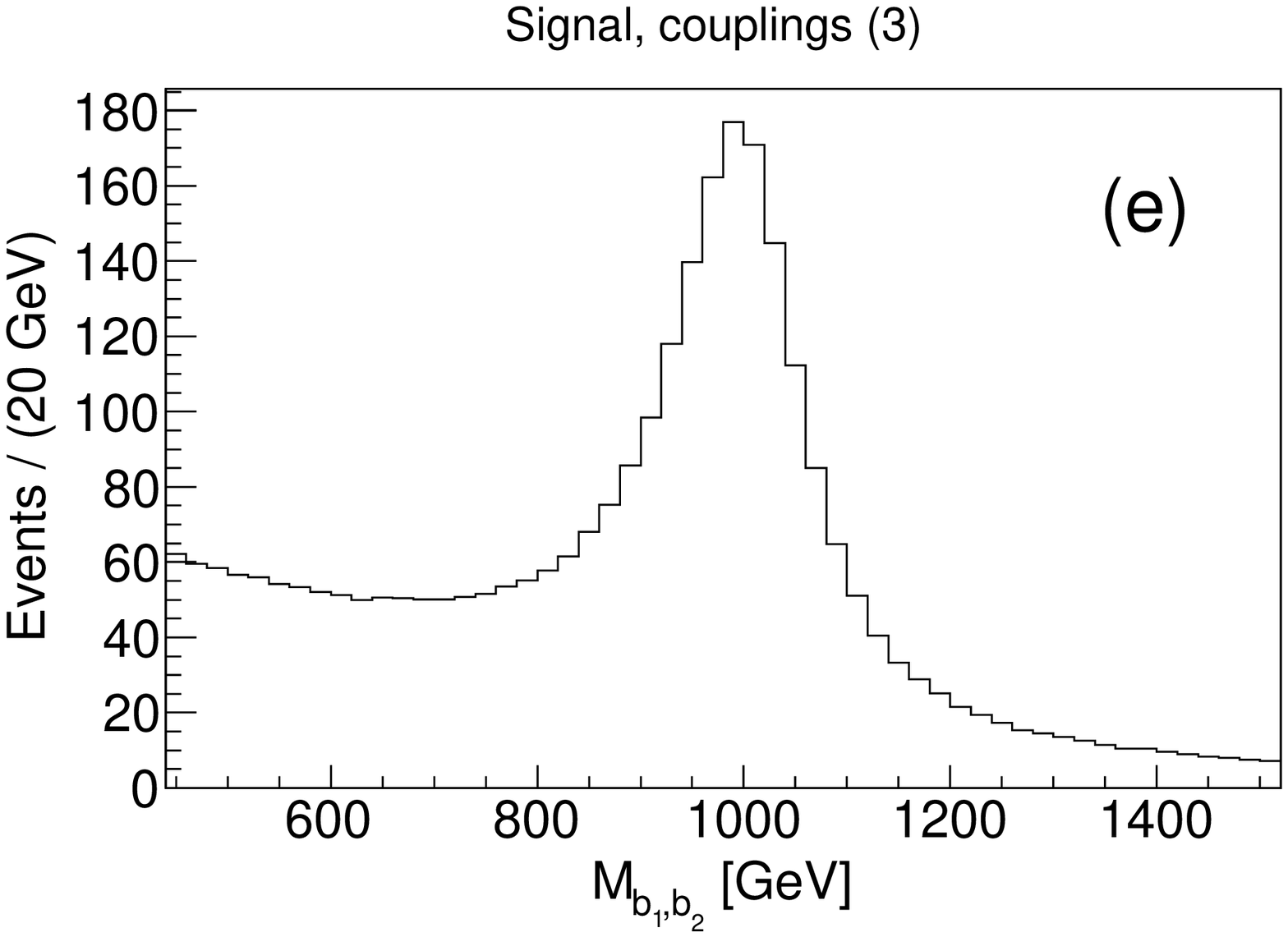}
  & 
  \epsfxsize=5.9cm
  \epsfbox{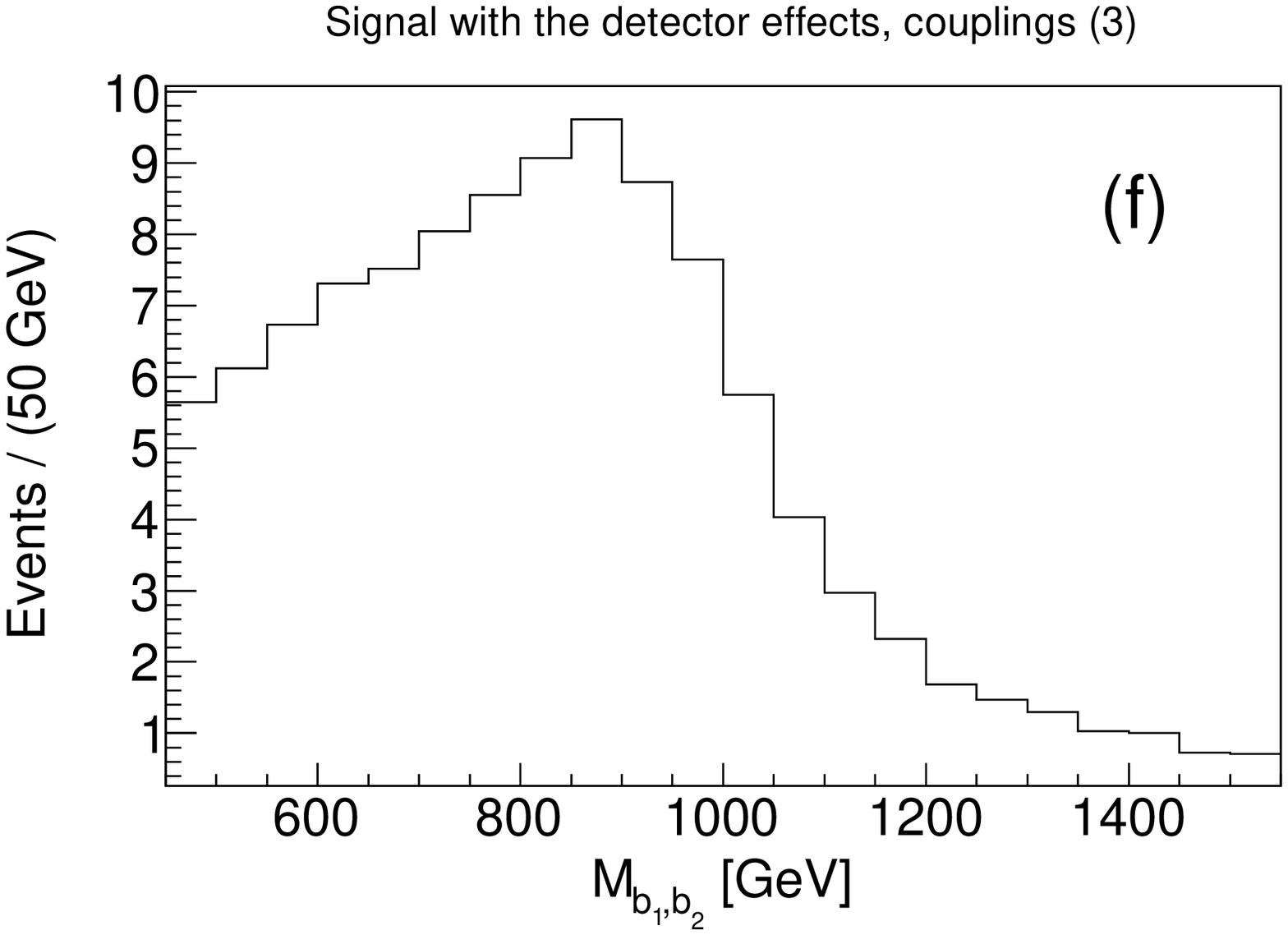} \\
  \epsfxsize=5.9cm
  \epsfbox{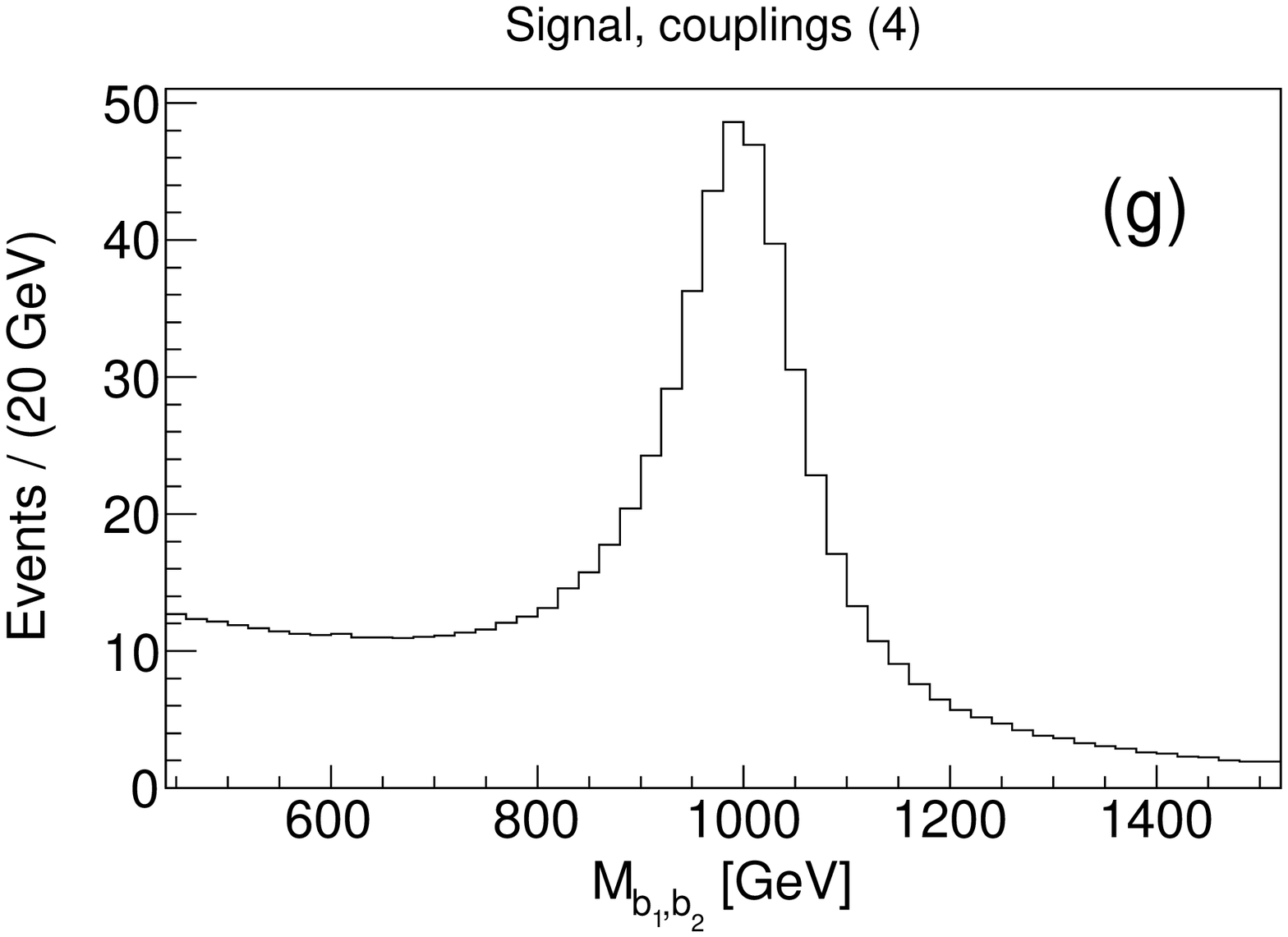}
  &
  \epsfxsize=5.9cm
  \epsfbox{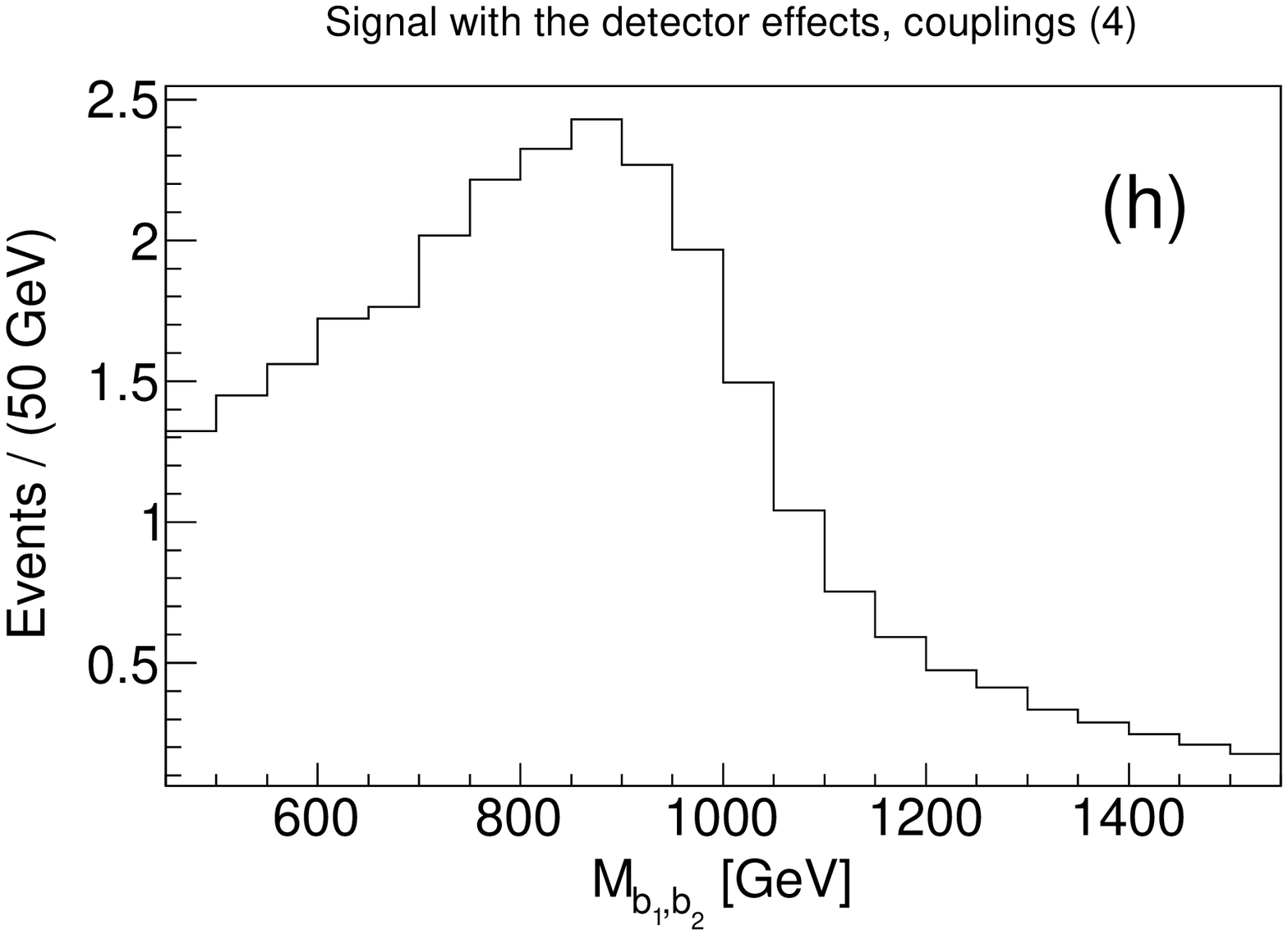} \\
 \end{array}
\end{eqnarray*}
\caption{
The invariant mass distribution of the $b_1b_2$ pairs of $b$-jets with the highest transverse momenta for the signal 
 process $pp\xrightarrow{g_{KK}^{(1)}} 3b$ without 
 ((a), (c), (e), and (g)) and with ((b), (d), (f), and (h)) the simulated effects 
 of the ATLAS detector and the Selection criterion~1 (with 
 $M_{b_1b_2}^{min}$ = 450 GeV). $M_{\gkk} = 1$~TeV was assumed and 
 four scenarios with couplings (\ref{couplings1})--(\ref{couplings4}) 
 were studied (marked as (1)--(4), in the figure). The number 
 of events in the histogram is scaled to the integrated luminosity of 10 
 fb$^{-1}$ for $pp$ collisions at $\sqrt{s}=14$ TeV.  
}
  \label{signal}
\end{center}
\end{figure}

\begin{figure}[htb]
\begin{center}
\begin{eqnarray*}
 \begin{array}{cc}
  \epsfxsize=8cm
  \epsfbox{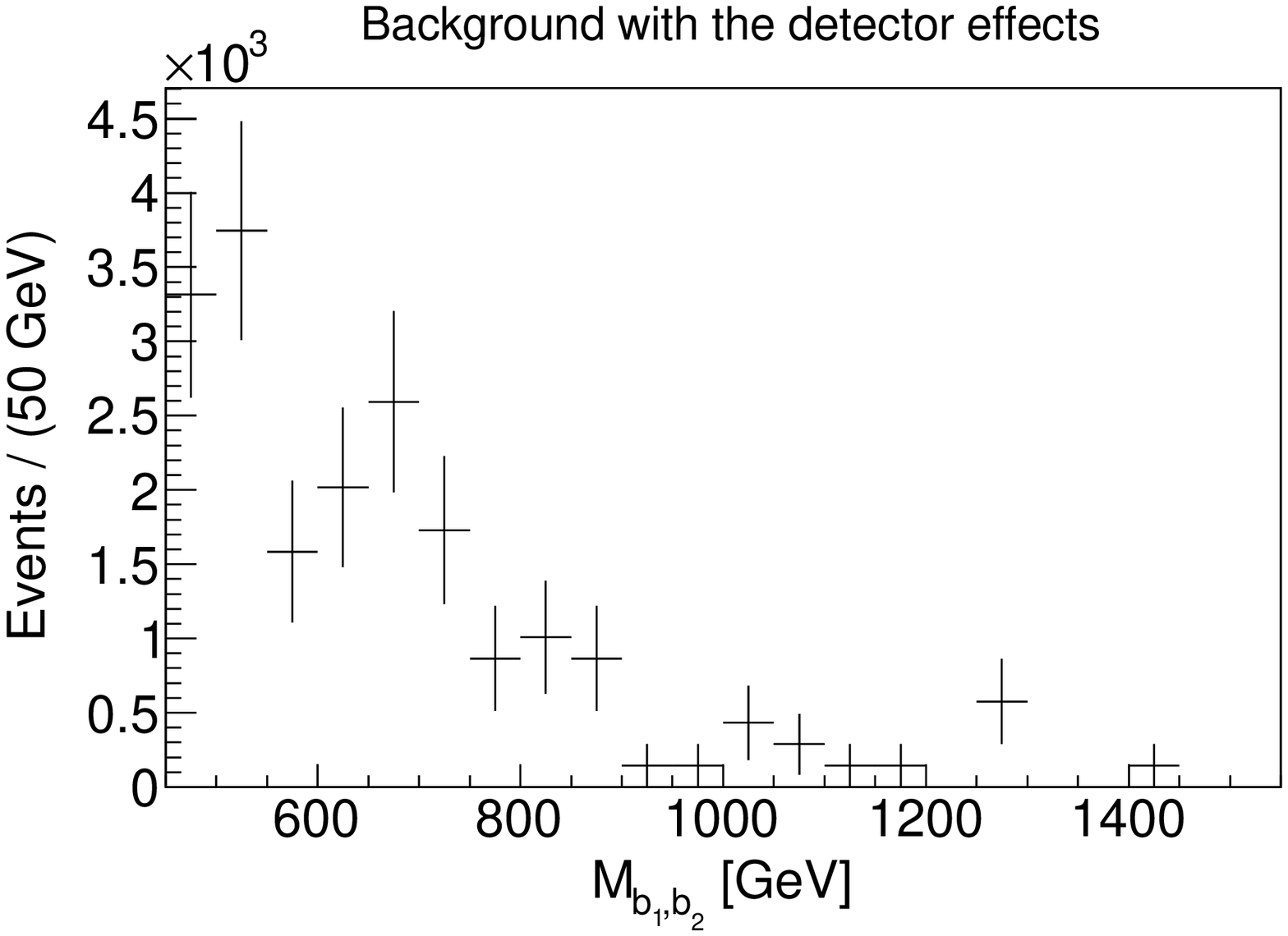}
  &
  \epsfxsize=8cm
  \epsfbox{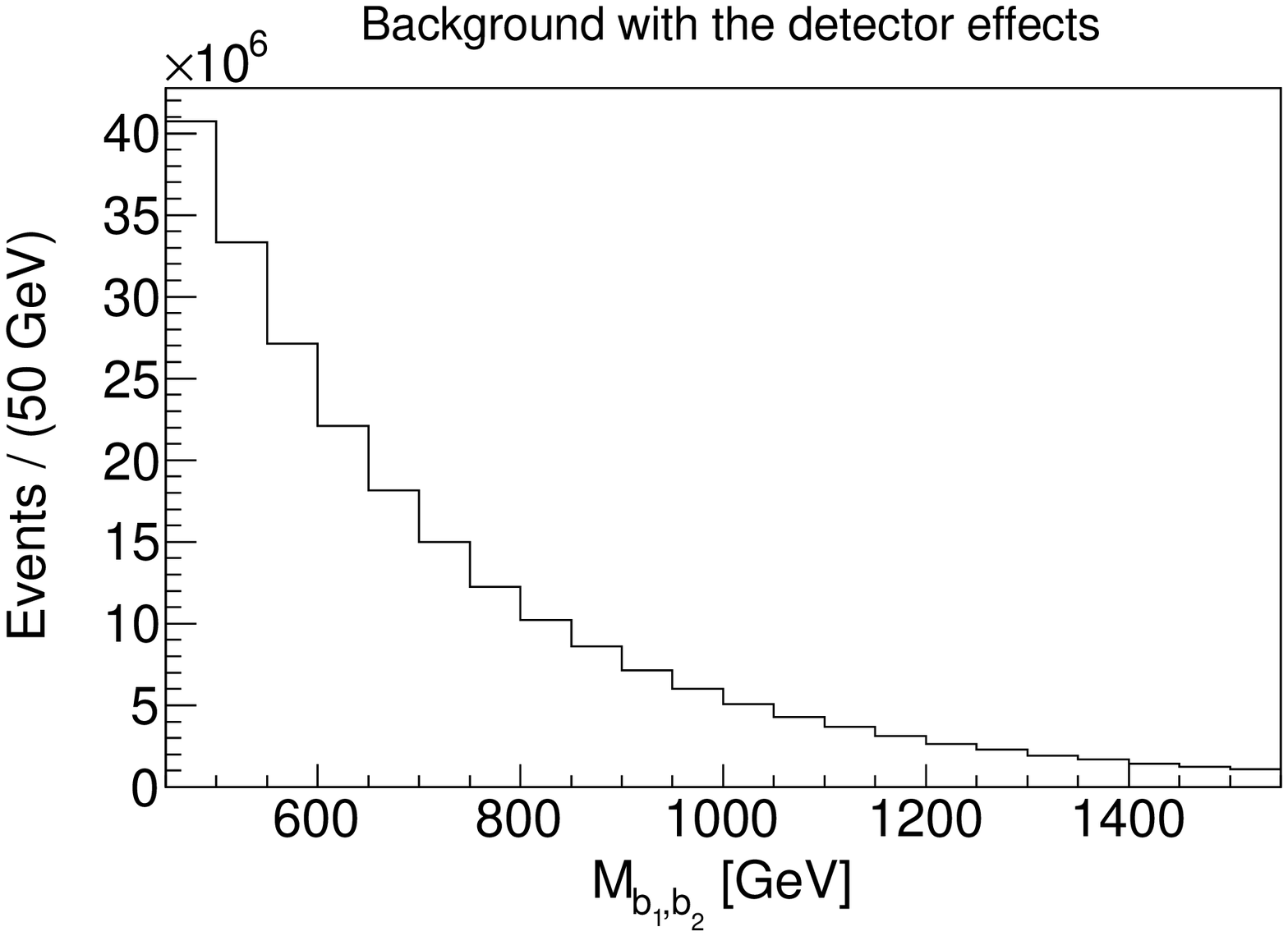} \\
 \end{array}
\end{eqnarray*}
\caption{
The invariant mass distribution of the selected jet pairs for the background processes with the simulated effects of the ATLAS detector and the Selection criterion 1 on the left side and the Selection criterion 2 on the right side. $M_{b_1b_2}^{min}$ in both selection criteria was chosen to be 450 GeV. The number of events in the histogram is scaled to the integrated luminosity of 10 fb$^{-1}$ for $pp$ collisions at $\sqrt{s}=14$ TeV. 
}
  \label{background}
\end{center}
\end{figure}

 We simulated the signal process for the four scenarios with the couplings
 (\ref{couplings1})--(\ref{couplings4}) and for the masses of the KK gluon
 between 1~TeV and 2.5~TeV. For the analysis, samples of $10^6$ signal
 events were used. Assuming the integrated luminosity of 10 fb$^{-1}$
 per year (during a low luminosity LHC run), it corresponds to the data
 collected during the period with the length at least 2 years, 
depending on the $\gkk$ mass and couplings. In the 
 Fig. \ref{signal}, distributions of a $b_1b_2$ invariant mass without
 and with the simulated detector effects and the Selection criterion 1 for
 $M_{\gkk} = 1$~TeV and couplings
 (\ref{couplings1})--(\ref{couplings4}) are presented. All plots are
 scaled to the integrated luminosity of 10 fb$^{-1}$.  

For the presented selection criteria, the most important background processes are the QCD production of three partons  (quarks, antiquarks and gluons). In the analysis, we used $100\times 10^6$ simulated background events. Due to an extremely high cross section, it corresponds to the data collected only during $7\times 10^{-3}$ year (assuming the integrated luminosity of 10 fb$^{-1}$ per year). In the Fig. \ref{background}, the invariant mass distribution of the detected $b$-jets (or objects supposed to be $b$-jets) is plotted. On the left side, the events were selected 
by using the Selection criterion 1 
(where the $b$-tagging was used). Due to low statistics and rescaling the histogram, the error bars for each bin were plotted. 
The plot on the right side was performed by using the Selection criterion 2 (where the $b$-tagging was not used). Applying 
the $b$-tagging reduces background events dramatically, however 
the low $b$-tagging efficiency can influence signal negatively, too: A number of observed signals are reduced.

\begin{table}[hbt]
\centering
\begin{tabular}{c c c c c c c c c}
\hline
\multirow{2}{*}{Scenario} & \multirow{2}{*}{$\frac{g^{(1)}_{\rm light}}{g_4}$} & \multirow{2}{*}{$\frac{g^{(1)}_{Q_3}}{g_4}$} & \multirow{2}{*}{$\frac{g^{(1)}_b}{g_4}$} & \multirow{2}{*}{$\frac{g^{(1)}_t}{g_4}$} & $M_{\gkk}$ & $M_{b_1b_2}^{min}$ & $S/\sqrt{B}$    & $S/\sqrt{B}$ \\
       &                      &                          &                       &                       & [TeV]            & [GeV]          & for 10 fb$^{-1}$ & for 100 fb$^{-1}$ \\ 
\hline
\hline
\multirow{4}{*}{(1)} & \multirow{4}{*}{0} & \multirow{4}{*}{4}  & \multirow{4}{*}{4}  & \multirow{4}{*}{4}    & 1.0    & 550   & $62 \pm 3$        & $206 \pm 10$ \\
%  &                           &                          &                       &                       			  			         & 1.2    & 900   & $32 \pm 4$       & $107 \pm 13$ \\
      &                          &                          &                       &                       						         & 1.5    & 900   & $13 \pm 2$       & $43 \pm 7$ \\ 
      &                          &                          &                       &                                                                          	         & 2.0    & 1300 & $4 \pm 1$         & $13 \pm 3$ \\ 
      &                          &                          &                       &                       						         & 2.5    & 1450 & $1.5 \pm 0.7$  & $5 \pm 2$ \\ 
\hline
\multirow{4}{*}{(2)} & \multirow{4}{*}{0} & \multirow{4}{*}{1} & \multirow{4}{*}{4}  & \multirow{4}{*}{4}     & 1.0     & 720   & $36 \pm 3$       & $120 \pm 10$ \\
%  &                          &                          &                       &                      					                   & 1.2     & 900   & $18 \pm 2$       & $60 \pm 7$ \\
      &                          &                          &                       &                       					                   & 1.5     & 1300 & $8 \pm 3$         & $27 \pm 10$ \\ 
      &                          &                          &                       &                       					                   & 2.0     & 1450 & $2 \pm 1$         & $7 \pm 3$ \\ 
      &                          &                          &                       &                       					                   & 2.5     & 1450 & $0.7 \pm 0.4$  & $2 \pm 1$ \\ 
\hline
\multirow{1}{*}{(3)} & \multirow{1}{*}{0} & \multirow{1}{*}{1} & \multirow{1}{*}{1} & \multirow{1}{*}{4}     & 1.0      & 730  & $1.04 \pm 0.09$ & $3.1 \pm 0.3$ \\
\hline
\multirow{1}{*}{(4)} & \multirow{1}{*}{0} & \multirow{1}{*}{1} & \multirow{1}{*}{0} & \multirow{1}{*}{4}     & 1.0      & 750  & $0.26 \pm 0.02$ & $0.9 \pm 0.07$ \\
\hline
\end{tabular}
\caption{The statistical significance $S/\sqrt{B}$ of our model for various values of $M_{\gkk}$ and couplings estimated for 10 fb$^{-1}$ and 100 fb$^{-1}$. The Selection criterion 1 using the $b$-tagging was used for the selection of the events. The presented errors correspond to the statistical errors related to our Monte-Carlo simulations.}
\label{TabSignificance1}
\end{table}

\begin{table}[hbt]
\centering
\begin{tabular}{c c c c c c c c c}
\hline
\multirow{2}{*}{Scenario} & \multirow{2}{*}{$\frac{g^{(1)}_{\rm light}}{g_4}$} & \multirow{2}{*}{$\frac{g^{(1)}_{Q_3}}{g_4}$} & \multirow{2}{*}{$\frac{g^{(1)}_b}{g_4}$} & \multirow{2}{*}{$\frac{g^{(1)}_t}{g_4}$} & $M_{\gkk}$ & $M_{b_1b_2}^{min}$ & $S/\sqrt{B}$    & $S/\sqrt{B}$ \\
     &                        &                          &                       &                       & [TeV]            & [GeV]          & for 10 fb$^{-1}$ & for 100 fb$^{-1}$ \\ 
\hline
\hline
\multirow{4}{*}{(1)} & \multirow{3}{*}{0} & \multirow{3}{*}{4}  & \multirow{3}{*}{4}    & \multirow{3}{*}{4}    & 1.0    & 500   & $9.23 \pm 0.02$     & $30.74 \pm 0.07$ \\
    &                          &                          &                       &                       					                     & 1.2    & 550   & $4.323 \pm 0.009$ & $14.39 \pm 0.03$ \\
    &                          &                          &                       &                       					                     & 1.5    & 550   & $1.650 \pm 0.003$ & $5.45 \pm 0.01$ \\ 
\hline
\multirow{3}{*}{(2)} & \multirow{3}{*}{0} & \multirow{3}{*}{1}  & \multirow{3}{*}{4}    & \multirow{3}{*}{4}    & 1.0     & 600   & $4.975 \pm 0.009$   & $16.57 \pm 0.03$ \\
     &                        &                          &                        &                       					                     & 1.2     & 750   & $2.203 \pm 0.005$   & $7.34   \pm 0.02$ \\
     &                        &                          &                        &                       					                     & 1.5     & 900   & $0.767 \pm 0.002$   & $2.554   \pm 0.007$ \\
\hline
\multirow{1}{*}{(3)} & \multirow{1}{*}{0} & \multirow{1}{*}{1} & \multirow{1}{*}{1} & \multirow{1}{*}{4}       & 1.0      & 650  & $0.1339 \pm 0.0002$ & $0.4017 \pm 0.0006$ \\
\hline
\multirow{1}{*}{(4)} & \multirow{1}{*}{0} & \multirow{1}{*}{1}  & \multirow{1}{*}{0}    & \multirow{1}{*}{4}   & 1.0      & 700  & $0.035 \pm 0.001$ & $0.117 \pm 0.003$ \\
\hline
\end{tabular}
\caption{The statistical significance $S/\sqrt{B}$ of our model for various values of $M_{\gkk}$ and couplings estimated for 10 fb$^{-1}$ and 100 fb$^{-1}$. The Selection criterion 2 without using the $b$-tagging was used for the selection of the events. The presented errors correspond to the statistical errors related to our Monte-Carlo simulations.}
\label{TabSignificance2}
\end{table}

As a signature of new physics, we use the number of selected events. For the integrated luminosity of 10~fb$^{-1}$ and 100~fb$^{-1}$, we estimated the number of expected observed signal and background events ($S$ and $B$) and the statistical significance $S/\sqrt{B}$. The significance of the deviation from the SM is proportional to the square root of the integrated luminosity. Therefore, it is easy to recompute the results for the higher integrated luminosity. We studied effects of variation of $M_{b\bar{b}}^{min}$ on the statistical significance. In the presented results, we use the value of $M_{b\bar{b}}^{min}$, for which the statistical significance $S/\sqrt{B}$ is maximal.

In the Tab.~\ref{TabSignificance1}, we present the statistical significance $S/\sqrt{B}$ of our model for various values of
$M_{\gkk}$ and couplings.  The Selection criterion 1 with $b$-tagging was used for the selection of the events. 
As expected, the deviation from the SM is strongly dependent on the coupling of a right-handed $b$ quark to a KK gluon. 
For the first set of couplings (\ref{couplings1}), the effects of KK gluons could be observable with the significance at least $4\sigma$ for the mass of a KK gluon up to 2~TeV and the integrated luminosity of 10 fb$^{-1}$ or for the mass of a KK gluon up to 2.5 TeV and the integrated luminosity of 100 fb$^{-1}$. For the second set of couplings (\ref{couplings2}), the effects of KK gluons could be observable with the significance of $8\sigma$ for the mass of a KK gluon up to 1.5 TeV and the integrated luminosity of 10 fb$^{-1}$ or with the similar significance for the mass of a KK gluon up to 2 TeV and the integrated luminosity of 100 fb$^{-1}$.  
For the third set of couplings (\ref{couplings3}), the effects of KK gluons are observable only for the integrated luminosity of 100~fb$^{-1}$ and $M_{\gkk} = 1$~TeV  with the significance of $3\sigma$. 
Due to the extremely low cross-section of the signal process, for the fourth set of couplings (\ref{couplings4}) the effects of KK gluons are unobservable. 

In the Tab.~\ref{TabSignificance2}, we present analogous results for the simulations using Selection criterion~2. Due to not applying 
the 
$b$-tagging, higher number of events are selected and hence the statistical errors of the obtained 
significance 
are smaller than 
ones of the results using the Selection criterion 1 as
presented in the Tab.~\ref{TabSignificance1}. For the couplings (\ref{couplings1}), the effects of KK gluons could be observable with the significance at least $4\sigma$ for the mass of a KK gluon up to 1.2 TeV and the integrated luminosity of 10 fb$^{-1}$ or for the mass of a KK gluon up to 1.5 TeV and the integrated luminosity of 100 fb$^{-1}$. For the couplings (\ref{couplings2}), the effects of KK gluons could be observable with the significance at least $5\sigma$ for the mass of a KK gluon up to 1 TeV and the integrated luminosity of 10 fb$^{-1}$ or for the mass of a KK gluon up to 1.2 TeV and the integrated luminosity of 100 fb$^{-1}$. 
Even for the integrated luminosity of 100 fb$^{-1}$, the effects of a KK gluon for the third set of couplings (\ref{couplings3}) are unobservable. 
The effects of a KK gluon for the fourth set of couplings (\ref{couplings4}) are unobservable
as like the result when the Selection criterion 1 is used.

\sect{Summary}

We examined the possibility of observing the effects of the first excitation of a KK gluon at the LHC. We focused on the final states with three $b$-jets. In our simulations, we studied $pp$ interactions with the energy $\sqrt{s} = 14$~TeV. For the estimation of the detector effects, we supposed the ATLAS detector. We studied two kinds of selection criteria -- with and without use of $b$-tagging (Selection criteria 1 and 2). We studied four sets of couplings (\ref{couplings1})--(\ref{couplings4}) of a KK gluon to $b$ and $t$ quarks. As a signature of new physics, we used the number of selected signal ($S$) and background ($B$) events and the ratio $S/\sqrt{B}$, the statistical significance of the observed phenomenon. 
For the Selection criterion 1 and for the integrated luminosity of 100 fb$^{-1}$, the effects of a KK gluon will be observable with 
the significance at least 5$\sigma$ for the scenario (\ref{couplings1}) with the KK gluon mass up to 2.5 TeV and
for the scenario (\ref{couplings2}) with the KK gluon mass up to 2~TeV. 
Even for the integrated luminosity of 10~fb$^{-1}$, the effects of KK
gluon will be observable with the 
significance at least 5$\sigma$ for both 
the 
scenarios 
(\ref{couplings1}) and (\ref{couplings2}) 
for the KK gluon mass up to 1.5 TeV. 
For the scenario (\ref{couplings3}) with a smaller coupling of a KK gluon to a right-handed $b$ quark will be observable only for the high integrated luminosity of 100 fb$^{-1}$ with  the significance of $3\sigma$.
The effects of a KK gluon for the scenario (\ref{couplings4}) with a vanishing coupling of a KK gluon to a right-handed $b$ quark, the deviation from the SM will not be observable. 
As expected, the effects of a KK gluon will be observable with the higher significance for the Selection criterion 1. However, the signatures of a KK gluon will be evident even for the results applying the Selection criterion 2 not using the $b$-tagging, though for lower KK gluon masses. For the integrated luminosity of 100 fb$^{-1}$, the effects of a KK gluon will be observable with the
significance at least 5$\sigma$ for the couplings (\ref{couplings1}) and (\ref{couplings2}) and the KK gluon mass up to 1.2 TeV.

If we compare our presented results for three $b$ final states with ones
for $b\bar{b}$ final states \cite{ArChSmYo}, we can conclude that the channel with three $b$-jets are more sensitive to the effects of the
KK gluon. It can be explained with a tough selection requiring three $b$-jets in the final state. It causes a sufficient suppression of the QCD background. 
On the other hand, the cross section of 
the signal process $pp\rightarrow 3b$
is roughly comparable to the cross section of 
$pp \rightarrow b\bar{b}$.
Initial partons forming the final state with two $b$-jets are both $b$ quarks, whereas initial partons forming the final state with three $b$-jets are one $b$ quark and one gluon. The second configuration is more preferable 
since the distribution functions of a gluon enhances the cross section 
of $pp\rightarrow 3b$ 
integrating over the momentum.

\noindent
{\bf Acknowledgements}\\
The work of M.A. is supported in part by Grants-in-Aid for Scientific 
Research from the Ministry of Education, Culture, Sports, Science and Technology 
(No.25400280).
The work of M.A. and K.S. is supported in part by the Research Program
MSM6840770029 and by the project of International Cooperation ATLAS-CERN LG13009
of the Ministry of Education, Youth and Sports of the Czech Republic. 
The work of G.C.C is supported in part by Grants-in-Aid for Scientific 
Research from the Ministry of Education, Culture, Sports, Science and Technology 
(No.24104502) and from the Japan Society for the Promotion of Science (No.21244036).  
Numerical results published in this work were computed with the Supercluster 
of the Computing and Information Centre of the Czech Technical University in Prague.

\small{

\end{document}
\begin{thebibliography}{66}
%
\bibitem{RS}
  L.~Randall and R.~Sundrum,
  %``A Large mass hierarchy from a small extra dimension,''
  Phys.\ Rev.\ Lett.\  {\bf 83} (1999) 3370
  [{\tt arXiv:hep-ph/9905221}];
 % L.~Randall and R.~Sundrum,
  %``An Alternative to compactification,''
  Phys.\ Rev.\ Lett.\  {\bf 83} (1999) 4690
  [{\tt arXiv:hep-th/9906064}].
  %%CITATION = HEP-PH/9905221;%%
  
\bibitem{Davoudiasl:1999tf}
 H.~Davoudiasl, J.~L.~Hewett and T.~G.~Rizzo,
  %``Bulk gauge fields in the Randall-Sundrum model,''
  Phys.\ Lett.\ B {\bf 473} (2000) 43
  [{\tt arXiv:hep-ph/9911262}].
  %%CITATION = HEP-PH/9911262;%%

\bibitem{Davoudiasl:2000wi}
  H.~Davoudiasl, J.~L.~Hewett and T.~G.~Rizzo,
  %``Experimental probes of localized gravity: On and off the wall,''
  Phys.\ Rev.\ D {\bf 63} (2001) 075004
  [{\tt arXiv:hep-ph/0006041}].
  %%CITATION = PHRVA,D63,075004;%%

\bibitem{JuWe}
  S.~Jung and J.~D.~Wells,
  %``Low-scale warped extra dimension and its predilection for multiple top quarks,''
  JHEP {\bf 1011} (2010) 001
  [arXiv:1008.0870 [hep-ph]].
  %%CITATION = ARXIV:1008.0870;%

\bibitem{Agashe:2003zs}
  K.~Agashe, A.~Delgado, M.~J.~May and R.~Sundrum,
  %``RS1, custodial isospin and precision tests,''
  JHEP {\bf 0308} (2003) 050
  [{\tt arXiv:hep-ph/0308036}].  %%CITATION = JHEPA,0308,050;%%

\bibitem{HePeRi}
  J.~L.~Hewett, F.~J.~Petriello and T.~G.~Rizzo,
  %``Precision measurements and fermion geography in the Randall-Sundrum model revisited,''
  JHEP {\bf 0209} (2002) 030
  [{\tt arXiv:hep-ph/0203091}].
  %%CITATION = JHEPA,0209,030;%%

%\bibitem{GrNe}
%  Y.~Grossman, M.~Neubert and ,
%  %``Neutrino masses and mixings in nonfactorizable geometry,''
%  Phys.\ Lett.\ B {\bf 474} (2000) 361
%  [hep-ph/9912408].
%  %%CITATION = HEP-PH/9912408;%%

%\bibitem{Chang:1999nh}
%  S.~Chang, J.~Hisano, H.~Nakano, N.~Okada, M.~Yamaguchi and ,
%  %``Bulk standard model in the Randall-Sundrum background,''
%  Phys.\ Rev.\ D {\bf 62} (2000) 084025
%  [hep-ph/9912498].
%  %%CITATION = HEP-PH/9912498;%%

\bibitem{GuMaSr1}
  M.~Guchait, F.~Mahmoudi and K.~Sridhar,
  %``Tevatron constraint on the Kaluza-Klein gluon of the Bulk Randall-Sundrum model,''
  JHEP {\bf 0705} (2007) 103
  [{\tt arXiv:hep-ph/0703060 [hep-ph]}].
    
\bibitem{LiRaWa}
  B.~Lillie, L.~Randall and L.~-T.~Wang,
  %``The Bulk RS KK-gluon at the LHC,''
  JHEP {\bf 0709} (2007) 074
  [{\tt arXiv:hep-ph/0701166}].
  %%CITATION = HEP-PH/0701166;%%

\bibitem{Chang:2008vx} 
  W.~-F.~Chang, J.~N.~Ng and J.~M.~S.~Wu,
  %``Flavour Changing Neutral Current Constraints from Kaluza-Klein Gluons and Quark Mass Matrices in RS1,''
  Phys.\ Rev.\ D {\bf 79} (2009) 056007 
  [{\tt arXiv:0809.1390 [hep-ph]}].
  %%CITATION = ARXIV:0809.1390;%%

\bibitem{GuMaSr2}
  M.~Guchait, F.~Mahmoudi and K.~Sridhar,
  %``Associated production of a Kaluza-Klein excitation of a gluon with a t anti-t pair at the LHC,''
  Phys.\ Lett.\ B {\bf 666} (2008) 347
  [{\tt arXiv:0710.2234 [hep-ph]}].

\bibitem{CDF}
CDF collaboration, 
%"A search for massive gluon decaying to top pair in lepton+jet channel," 
CDF note 9164.

\bibitem{CMS}
CMS collaboration, 
%"A Search for Resonances in Semileptonic Top Pair Production at $\sqrt{s}= 7 TeV$," 
CMS PS TOP-11-009.

\bibitem{ATLAS}
ATLAS collaboration, 
%"A search for $t\bar{t}$ resonances in the lepton plus jets channel using 2.05 fb${}^{-1}$ of pp collisions at $\sqrt{s} = 7$ TeV," 
ATLAS-CONF-2012-029.

\bibitem{ArChSmYo}
M.~Arai, G-C.~Cho, K.~Smolek, and K.~Yoneyama,
 Phys.\ Rev.\ D {\bf 87} (2013) 016010
  [{\tt arXiv:1211.7006 [hep-ph]}].

\bibitem{AgBeKrPeVi} 
  K.~Agashe, A.~Belyaev, T.~Krupovnickas, G.~Perez and J.~Virzi,
  %``LHC Signals from Warped Extra Dimensions,''
  Phys.\ Rev.\ D {\bf 77} (2008) 015003
  [{\tt arXiv:hep-ph/0612015}].

\bibitem{GhPo}
  T.~Gherghetta and A.~Pomarol,
  %``Bulk fields and supersymmetry in a slice of AdS,''
  Nucl.\ Phys.\ B {\bf 586} (2000) 141
  [{\tt arXiv:hep-ph/0003129}].

\bibitem{Allanach:2009vz} 
  B.~C.~Allanach, F.~Mahmoudi, J.~P.~Skittrall, K.~Sridhar,
  %``Gluon-initiated production of a Kaluza-Klein gluon in a Bulk Randall-Sundrum model,''
  JHEP {\bf 1003} (2010) 014 
  [{\tt arXiv:0910.1350 [hep-ph]}].

\bibitem{Cho:2009rj}
  G.~-C.~Cho and Y.~Kanehata,
  %``Kaluza-Klein gluon and b-jet forward-backward asymmetry,''
  Phys.\ Lett.\ B {\bf 694} (2010) 134
  [{\tt arXiv:0902.3322 [hep-ph]}].

\bibitem{MadGraph}
 % MadGraph
J.~Alwall, M.~Herquet, F.~Maltoni, O.~Mattelaer and T.~Stelzer, JHEP {\bf 1106} (2011) 128
[{\tt arXiv:1106.0522 [hep-ph]}].

\bibitem{Christensen:2008py}
% FeynRules 
  N.~D.~Christensen and C.~Duhr,
  %``FeynRules - Feynman rules made easy,''
  Comput.\ Phys.\ Commun.\  {\bf 180} (2009) 1614
  [{\tt arXiv:0806.4194 [hep-ph]}].
  %%CITATION = ARXIV:0806.4194;%%
  %213 citations counted in INSPIRE as of 09 Jul 2013

\bibitem{Pythia6}
  % Pythia 6.4 Physics and Manual
  T.~Sj\"{o}strand, S.~Mrenna and P.~Skands, JHEP {\bf 05} (2006) 026
  [{\tt arXiv:hep-ph/0603175}].

\bibitem{Pythia8}
  % A Brief Introduction to Pythia 8.1
  T.~Sj\"{o}strand, S.~Mrenna and P.~Skands, Comput. Phys. Comm. {\bf 178} (2008) 852
  [{\tt arXiv:0710.3820 [hep-ph]}].

\bibitem{CTEQ6L1}
  % CTEQ6L1 PDF
  J.~Pumplin, D.~R.~Stump, J.~Huston, H.~L.~Lai, P.~M.~Nadolsky and W.~K.~Tung, JHEP {\bf 0207} (2002) 012
  [{\tt arXiv:hep-ph/0201195}].

\bibitem{Delphes}
  % Delphes
  S.~Ovyn, X.~Rouby, and V.~Lemaitre, 
  %{\it Delphes, a framework for fast simulation of a generic collider experiment} 
  [{\tt arXiv:0903.2225 [hep-ph]}].

\bibitem{FastJet1}
  M.~Cacciari, G.~P.~Salam and G.~Soyez, Eur. Phys. J. C {\bf 72} (2012) 1896 
  [{\tt arXiv:1111.6097 [hep-ph]}].

\bibitem{FastJet2}
  M.~Cacciari, G.~P.~Salam, Phys. Lett. B {\bf 641} (2006) 57 
  [{\tt arXiv:hep-ph/0512210}].

\bibitem{kT_algorithm}
  M.~Cacciari, G.~P.~Salam and G.~Soyez, JHEP {\bf 0804} (2008) 063
  [{\tt arXiv:0802.1189 [hep-ph]}].


\end{thebibliography}
